\documentclass{aastex}
\usepackage{graphicx}
\newcommand{\kms}{km\,s$^{-1}$}
\begin{document}   
\title{Active star formation in N11B Nebula in the Large Magellanic Cloud:
        a sequential star formation scenario confirmed
        \footnotemark~~\footnotemark}

\footnotetext[1]{Based in part on observations with the NASA/ESA Hubble Space Telescope obtained from the archive at the Space Telescope Science Institute, which is operated by the Association of Universities for Research in Astronomy, Inc., under NASA contract NAS5-26555}
\footnotetext[2]{Based in part on observations obtained at European Southern Observatory, La Silla, Chile}

\author{Rodolfo H. Barb\'a\altaffilmark{{3}}}
\affil{Observatorio Astron\'omico La Plata, Paseo del Bosque s/n, B1900FWA, La Plata, Argentina}
\authoremail{rbarba@fcaglp.edu.ar}

\author{M\'onica Rubio\altaffilmark{{4}}}
\affil{Departamento de Astronom\'{\i}a, Universidad de Chile, Casilla 36-D, Santiago, Chile}
\authoremail{mrubio@das.uchile.cl}

\author{Miguel R. Roth}
\affil{Las Campanas Observatory, The Observatories, Carnegie Institution of Washington, Casilla 601, La Serena, Chile}
\authoremail{miguel@lco.cl}

\author{{Jorge Garc\'{\i}a}}

\affil{Gemini Observatory Southern Operations Center, c/o AURA, Inc., 
Casilla 603, La Serena, Chile}
\authoremail{jgarcia@gemini.edu}

\altaffiltext{3}{Member of Carrera del Investigador Cient\'{\i}fico, CONICET, Argentina}    
\altaffiltext{4}{Visiting Astronomer, Las Campanas Observatory, Chile}

\received{}
\revised{}

\shorttitle{Active star formation in the N11B Nebula}
\shortauthors{Barb\'a et al.}

\begin{abstract}   
The second largest \ion{H}{2} region in the Large Magellanic Cloud, N11B has 
been surveyed in the near IR. 
We present $JHKs$ images of the N11B nebula. 
These images are combined with CO$(1\rightarrow0)$ emission line data and 
with archival NTT and HST/WFPC2 optical images to address the star formation 
activity of the region.
IR photometry of all the IR sources detected is given.
We confirm that a second generation of stars is currently forming in the N11B 
region. 
Our IR images show the presence of several bright IR sources which 
appear located towards the molecular cloud as seen from the CO emission in 
the area.
Several of these sources show IR colours with YSO characteristics and they are
prime candidates to be intermediate-mass Herbig Ae/Be stars.
For the first time an extragalactic methanol maser is directly associated 
with IR sources embedded in a molecular core. 
Two IR sources are found at $2''$ (0.5 pc) of the methanol maser reported 
position. 
Additionally, we present the association of the N11A compact \ion{H}{2} 
region to the molecular gas where we find that the young massive O stars have 
eroded a cavity in the parental molecular cloud, typical of a champagne flow. 
The N11 region turns out to be a very good laboratory for studying the 
interaction of winds, UV radiation and molecular gas. 
Several photodissociation regions are found.

\end{abstract}


\section{Introduction}

Giant \ion{H}{2} regions are relatively scarce objects in the Local Group. 
In these regions we expect to find a broad spectrum of coeval phenomena 
related to massive stars.
The action of stellar winds from the massive stars and the supernova 
explosions pushes and destroys the natal molecular cloud, but also, helps to 
trigger the formation of a new generation of massive stars. 
At present, there is morphological evidence showing that a former massive 
star generation could produce a new one in the peripheral molecular clouds, 
but there is no quantitative evidence of how different generations are related 
in a giant \ion{H}{2} region.
This quantitative evidence is very hard to obtain, because it requires
the characterisation of the stellar content embedded in the \ion{H}{2} 
region and the physical condition of the gas in the molecular cloud.

The sequential star formation scenario proposed by Elmegreen \& Lada
(1977, see also Elmegreen 1998), has a few clear galactic examples where
it is possible to see that the direct action of massive stars in the
parental molecular cloud could be producing a new generation of stars.
Two such regions, the OB Association Ara OB1 (Arnal et al. 1987), and the 
Carina Nebula (Smith et al. 2000) have been proposed as leading cases in our 
Galaxy.

In general, in a massive star forming region we find a cavety of 
relatively low density
ionised gas produced by the action of several hot massive stars 
together with photodissociation regions (PDR) located on the surfaces of the 
molecular clouds facing the hot stars.  
Embedded IR sources and/or molecular cores could be direct evidence of an 
emerging  new generation of stars.
                            
In the Local Group, the formation and evolution of large superbubbles 
(Oey 1999) is a controversial issue. 
Evolved superbubbles are suggested to be produced by the direct action
of SN explosions in the interstellar medium which produce
shells of emitting gas (Wang \& Helfand 1991).  
In younger \ion{H}{2} superbubbles, the main bright nebular filaments are not 
shells but PDR interfaces between the \ion{H}{2} cavity and the surrounding 
molecular clouds surfaces. 
An example in an extragalactic environment is now clearly established: 
30 Doradus Nebula in the Large Magellanic Cloud (Rubio et al. 1998; 
Walborn et al. 1999a and references therein).

The N11 nebular complex (Henize 1956) or DEM34 (Davies, Elliot, \& 
Meaburn 1976) is the second largest \ion{H}{2} region in the LMC 
after 30 Doradus Nebula (Kennicutt \& Hodge 1986), lying at the opposite end 
of the LMC Bar. 
The complex consists of a huge bubble surrounded by nine distinct nebular 
entities (Rosado et al. 1996). 
These are easily distinguished in the spectacular 
H$\alpha$+[\ion{N}{2}] emission line image published by Walborn \& 
Parker (1992).
The OB association LH9 is located at the center of a cavity of about 
$80\times60$ pc (Lucke \& Hodge 1970; Parker et al. 1992).
This OB association is dominated by HD\,32228 (Radcliffe\,64, Sk\,$-66\,28$, 
Breysacher\,9), a massive compact cluster containing 
a WC4 star of about 3.5 Myr (Walborn et al. 1999b).
Surrounding LH9 are several younger OB associations embedded in 
dense nebular regions, two of which have O3 stars among their members: 
LH13 in N11C (Heydari-Malayeri et al. 2000) and LH10 in N11B (Parker et al. 
1992).
The latter authors propose an evolutionary link between the OB 
associations LH9 and LH10, where the star formation in LH10 could have been 
triggered by the evolution of massive stars in LH9. 
This suggestion was
based on the different initial mass function slopes found, 
significantly flatter for LH10 than for LH9, and the different apparent ages
derived for both associations. 
Walborn \& Parker (1992) noticed a remarkable analogy
in the structural morphology between N11 and 30 Doradus 
and proposed a two-stage starburst scenario to explain 
the morphological distribution of OB stars in both \ion{H}{2} regions. 
According to this scenario, an initial and centrally concentrated burst of 
stars may trigger a 
second burst in the peripherical molecular clouds about $2\times10^6$ years 
later. In N11, this process would be 2 million years older than in 30 Dor.

N11 is a large molecular cloud complex composed by at least 29 separate 
molecular clouds (Israel et al. 2002, Israel \& de Graauw 1991),  
directly associated with the ionised gas (Rosado et al. 1996). 
The CO emission is concentrated in distinctive peaks and correlated 
with the H$\alpha$ and FIR dust emissions, suggesting that the star-formation 
activity is distributed in the ringlike structure of N11 
(Caldwell \& Kutner 1996).

Thus, the N11 region is a very good candidate for the study of the sequential 
star formation processes and for the determination of how different complex 
elements are 
related in the ongoing star generation embedded in the nebular ring. 

In this paper we present new near-infrared images and CO observations of the 
N11B nebula which show a striking relationship between IR sources 
and the molecular cloud. In addition, we compare our observations with 
optical nebular emission archival images obtained with the New Technology 
Telescope (NTT) and the Hubble Space Telescope (HST) yielding further 
evidence of  the ongoing star forming activity in this region. 

\section{Observations}

\subsection{Millimeter}

Millimeter observations of N11 were done in the $^{12}$CO(1$\rightarrow$0) as 
part of the Swedish-ESO Submillimeter Telescope (SEST) Key Programme: 
``CO in the Magellanic Clouds'' between years 1988-94, using the SEST 
radiotelescope (ESO, Chile)\footnote{The Swedish-ESO Submillimeter Telescope
(SEST) is operated jointly by the European Southern Observatory (ESO) and the 
Swedish Science Research Council (NFR)}. 
These observations were done with a FWHP=$45''$ resolution (Israel \& de 
Graauw 1991; Israel et al. 2002), and the molecular clouds of the complex were 
fully mapped with a $20''$ grid spacing. The observations showed the presence 
of 29 individual molecular clouds which appeared well correlated with 
H$\alpha$ and FIR peaks. 
For this work we have selected a subset of the Key Programme CO observations, 
which include the regions N11A and N11B. The sampled area was $7'\times 3'$ 
with a grid spacing of $20''$ (5\,pc at the LMC distance).

The data were reduced using the millimeter IRAM software package. The spectra
were smoothed to a velocity resolution of 0.45 \kms. Integrated contour maps
of the region were produced and used in the following sections to
superimpose the molecular clouds on the NIR and optical nebular images.

\subsection{Infrared images}

The $J$, $H$ and $Kshort$ ($Ks$) images of the OB association LH10 in N11B,
centered around the star PGMW\,3070\footnote{PGMW star numbers are from 
Parker et al. 1992} were obtained on 28 January 1993, using the near-IR 
camera {\em IRCAM}, attached to the 2.5-m Du Pont Telescope at Las Campanas 
Observatory (LCO), Chile. {\em IRCAM} is equipped with a NICMOS III 
$256\times256$ array (Persson et al. 1992) and the pixel scale was
$0\farcs35$\,px$^{-1}$. 
The seeing during  the observations was typically of $1''$ giving an optimum 
sampling. Nine position tiles with a separation of $30''$ were observed in 
each filter. 
Each tile was formed of a sequence of five exposures of 40 sec each, and these 
were combined to obtain, a total on source exposure time of 800 sec in each 
filter.   
Sky frames were taken at one degree to the North of N11B, showing few stars 
and no extended nebulosities. 
In each filter, the images were median-averaged after subtraction of dark, 
and sky, and then flat-fielded using IRAF\footnote{IRAF is distributed 
by the National Optical Astronomy Observatories, which is operated by the 
Association of Universities for Research in Astronomy, Inc., under contract 
to the National Science Foundation.} routines. 
Several standard stars extracted from Persson et al. (1998) were observed 
during the night; their images were also dark-subtracted and and flat-fielded. 
All combined images of the LH10 region were registered with respect to the 
$Ks$ image by means of several common stars, and the final area covered  
resulted to be of $110''\times110''$ for each filter.

\subsection{Optical}

A 120 sec exposure time image obtained with EMMI Camera attached to the 
New Technology Telescope (NTT) at the European Southern Observatory, 
using the filter \#589 [\ion{O}{3}], was retrieved from the ESO Archive 
Facility. 
The image was obtained on Dec. 16, 1995 by J. Danziger using a 2K$\times$2K 
CCD (scale of $0\farcs266$\,px$^{-1}$) on very good seeing conditions.
The image was processed as follows: it was flat-fielded using flat images 
obtained during the same day, but due to the unavailability of bias images, 
the overscan region was used instead to remove the  bias level. Few cosmic 
rays were detected by visual inspection of the image and these were 
``cleaned'' by linear interpolation with neighbour pixels.

Archival images obtained with the HST/WFPC2 were used to investigate the 
morphological relationships between IR sources and stellar objects or gaseous 
structures in the OB association LH10.  
These WFPC2 datasets were obtained in May 12, 1999 (Proposal 6698, 
PI Y.-H. Chu), using narrowband filters F502N, corresponding to [\ion{O}{3}] 
5007\AA\ (datasets u3me0103r and u3me0104r, 1200 sec of exposure time) 
and F656N, 
corresponding to H$\alpha$ (datasets u3me0101r and u3me01012r, 1000 sec of 
exposure time).
These images were extracted from the HST Archive and  calibrated using the 
``on the fly'' calibration pipeline. Additional cosmic rays rejection, 
combining and mosaicing were done using IRAF/STSDAS software. The datasets are
the same as those reported by Naz\'e et al. (2001).

\section{Results}

\begin{figure}
\figurenum{1}
\epsscale{1.00}
\plotone{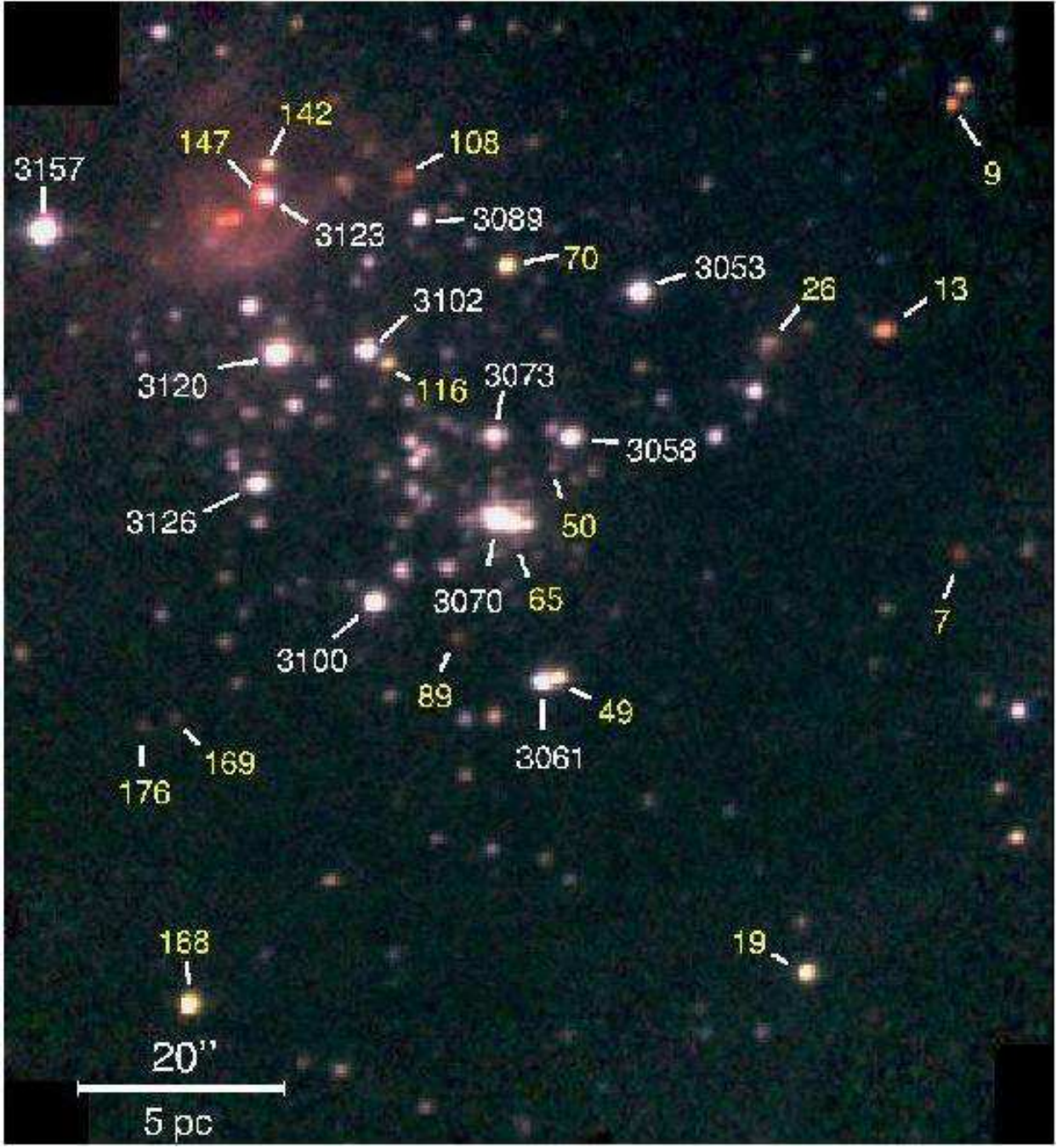}
\caption{Color composite of $J$, $H$ and $Ks$ (blue, green and red channels,
respectively) of part of the N11B nebula, around PGMW\,3070. North is up and
east to the left, as in all images in this paper. Numbers greater than
3000 denote optical stars from Parker et al. (1992, PGMW). Smaller numbers
(1--184) correspond to RGGB IR sources measured here.
\label{fig1-ir-n11b}}
\end{figure}

\begin{figure}
\figurenum{2}
\epsscale{1.00}
\plotone{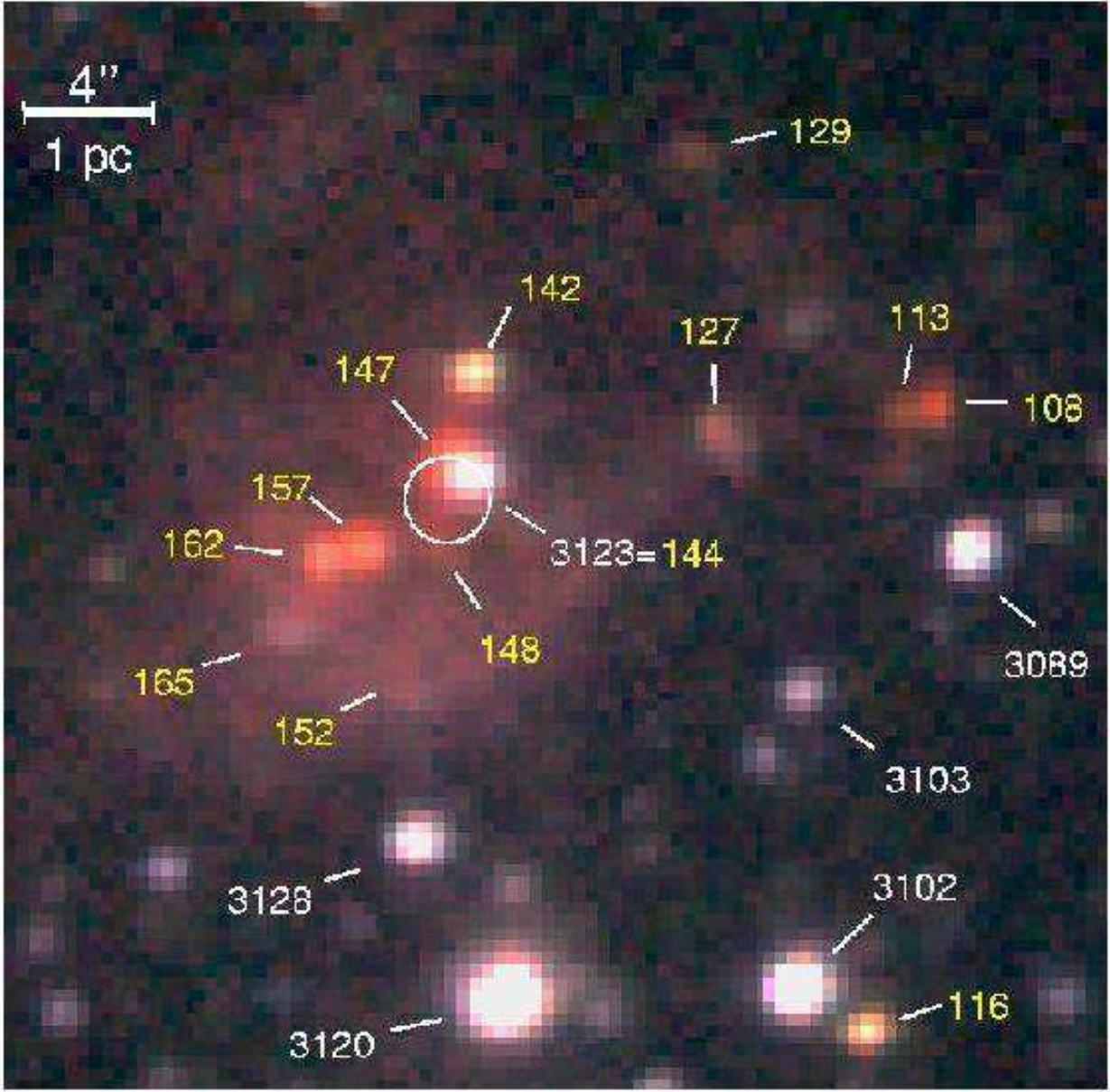}
\caption{
Close-up of the NE corner of Fig.~1 where several IR sources (yellow 
numbers) are located. The object designation is as in Fig.~1. 
The $1''$ circle shows the 
position of the methanol maser described in Section~4.2 .
\label{fig2-ir-n11b-zoom}}
\end{figure}

Figure~\ref{fig1-ir-n11b} is a {\em false-colour} image obtained  from a 
combination of $J$, $H$, and $Ks$ images as {\em blue}, {\em green}, 
and {\em red} channels, respectively. Several IR sources are found. These are
labelled in black numbers in the figure while those catalogued in the optical
are labelled with white numbers. The  image is centered on PGMW\,3070, 
the multiple star core of LH10 with an O6\,V spectral type, showing a cluster 
of blue stars with relatively low extinction. In the NE corner of the image, 
around the O8.5\,V star PGMW\,3123, a nebular feature appears with several 
embedded IR sources (shown in greater detail in 
Figure~\ref{fig2-ir-n11b-zoom}).
In this area, the WFPC2 and NTT images (discussed in following sections) 
show that this nebulosity is in fact a PDR at the interface between the 
molecular cloud and the \ion{H}{2} cavity fuelled by the UV photons produced 
by the hottest stars in LH10. 

\subsection{Photometry of infrared sources}

Point-spread function photometry was performed using IRAF/DAOPHOT-II 
software running in a Linux workstation. Stars were detected 
at a $5\sigma$ level above the mean background, and additional stars 
surrounded by nebulosities were included by eye. Three PSF star candidates 
were selected in each image avoiding nebular and crowded regions. The PSF was 
calculated using a {\it penny1} function and one look-up table. 
The final PSF photometry was made using a 3 pixel aperture (roughly the FWHM 
of the stars), and aperture corrections were estimated using the 
curve-of-growth method with the same PSF stars for each frame.  
False star detections were minimised by eye inspection of the images, mainly 
in the nebular region. Objects suspected to be false detections were rejected 
from the final catalogue. 
The photometric limits were determined plotting the star distribution per 
$0\fm5$ bin for each filter, and assuming that the peak of such distributions 
were magnitude limits. 
The peak of the distribution occurred about $1\fm-1\fm2$ brighter than the 
magnitude where the distribution fell to zero. Thus, we estimate that our 
$JHKs$ catalogue is complete to $J \lesssim 17.7$, $H \lesssim 17.3$, and 
$Ks \lesssim 17.0$.

Table~\ref{table1-ir-catalog} gives the photometry of all 184 IR sources 
detected in the field. 
Running number sources are in 
column 1 (BRRG numbers), 
column 2 and 3 are right ascension and declination (J2000),  
columns 4, 5, 6, 7, 8, 9 are $J$, $H$, $Ks$ magnitudes and their errors,
columns 10 and 11 are  $J-H$ and $H-Ks$ colours, 
and column 12 contains comments related to the identification 
of optical counterparts. 

\begin{figure}
\figurenum{3}
\epsscale{0.5}
\plotone{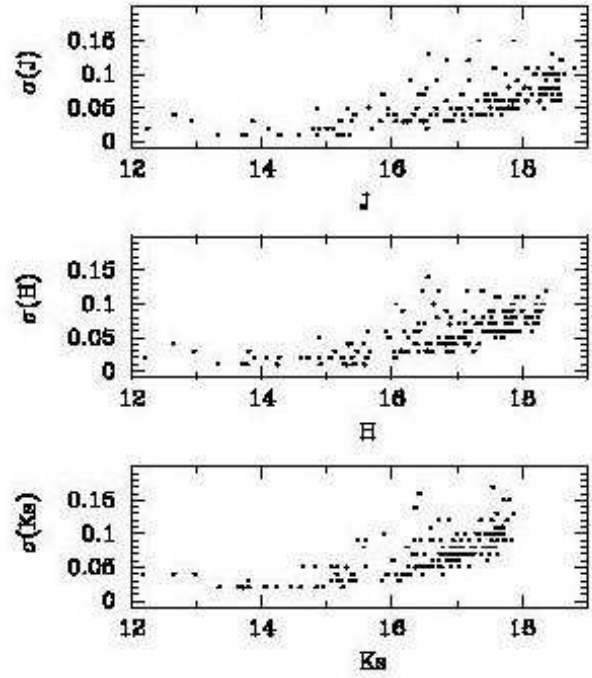}
\caption{Photometric errors derived from DAOPHOT as a function
of magnitude for each infrared filter.
\label{fig3-dao-errors}}
\end{figure}

Internal photometric errors as determined by DAOPHOT (including 
photon-counts statistics, NICMOS noise characteristics, and PSF  
fitting errors) are plotted in Figure~\ref{fig3-dao-errors}.
To check systematic photometric errors produced by aperture
corrections and standard zero-points, we compare un-crowded regions
with $Ks<14$ sources with those of the 2MASS point source catalogue
(Second Incremental Data Release, Cutri et al. 2000). 
Table~\ref{table2-comp-2MASS} lists the stars used for the comparison 
and the magnitude differences derived from our photometry and those of
the 2MASS point source catalogue.
A comparison of our photometry to that of the 2MASS catalogue
shows differences which are indicative that our photometry is 
systematically fainter than that obtained from the 2MASS. We find
a magnitude offset of about $0.10-0.15$ magnitudes 
between the LCO and 2MASS infrared photometry. 
We rechecked our sky-subtraction procedures, aperture corrections 
and zero-points to determine the origin of such differences 
and we have not found any error in our procedure.  
A possible explanation for these differences could be
a change in the observing conditions during the N11B observation
(for example, the passage of tiny atmospheric cirrus).
We decided to apply a correction to our final catalogue taking into 
account these differences with the 2MASS catalogue. 

Positions in equatorial coordinates of the individual sources were 
derived from the identification of seven Guide Star Catalogue (GSC) stars 
in the IR images.
Positions derived in this way show systematic differences with those 
derived from the default world coordinate system astrometric solution in the
HST/WFPC2 image headers. 
This offset in coordinates is not seen
when we compare our positions with those in other
catalogues such as Parker et al. (1992) and 2MASS. The offset difference 
found with the HST images is in the range of the absolute errors expected 
from the HST pointing system.
Table~\ref{table3-diff-coord} gives
the average offsets and errors found on the coordinates 
between our IR seven stars and those found in the 
different catalogues (GSC, Parker et al. 1992, 
2MASS) and the positions derived from WFPC2 images.
This offset coordinate difference must be known with the best attainable 
precision as we will compare the position of a methanol maser with that 
of detected IR sources and the HST/WFPC2 images (see Section~\ref{methanol}).

\subsection{Colour-Magnitude and Colour-Colour diagrams}
\label{photom_ir}

\begin{figure}
\figurenum{4}
\epsscale{0.60}
\includegraphics[angle=-90,scale=0.65]{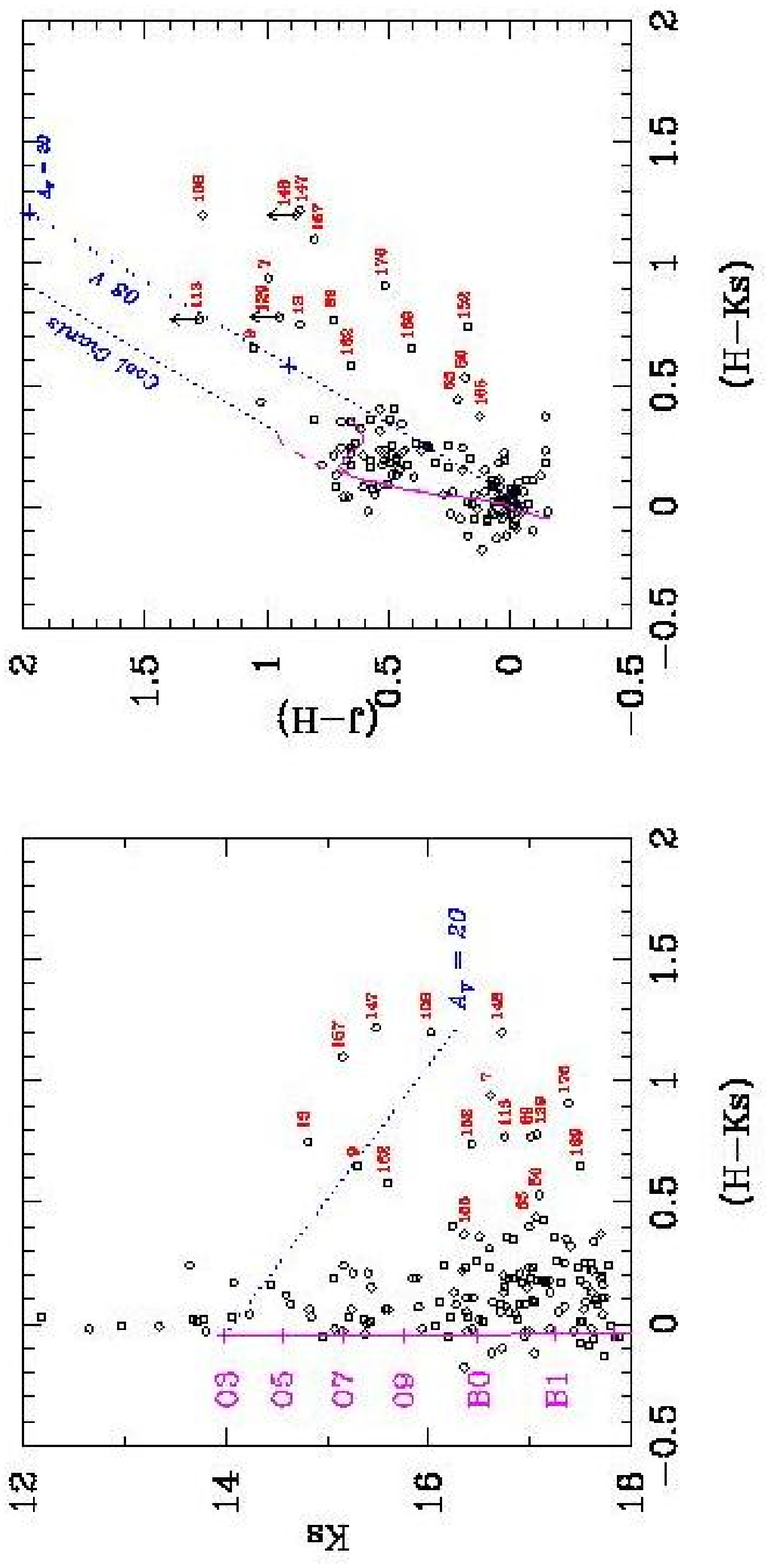}
\caption{(a) $Ks$ vs. $H-Ks$ color-magnitude diagram for the observed field
in N11B. The upper ZAMS between O3\,V and B1\,V, corresponding to a distance
modulus of 18.6, is indicated with a solid line. The reddening track for a 
normal O3\,V star is plotted with a dotted line and extends to $A_V=20$\,mag.
Numbers are from IR sources in Tables~2 and 3.
(b) $J-H$ vs. $H-Ks$ color-color diagram for the same objects. 
The main-sequence locus between O3\,V and M2\,V is indicated with a solid
line, while the cool-giant branch with a dash line. The reddening tracks for
normal O3\,V and cool-giant stars are plotted as dotted lines, with crosses
indicating $A_V=10$ and $20$ mag.
\label{fig4-diagrams}}
\end{figure}

Colour-Magnitude (CMD) and Colour-Colour diagrams (Fig~\ref{fig4-diagrams})
show a clear main-sequence between spectral types O3 and B0 on the ZAMS 
(our lower limit) spread out by low or moderate range of visual absorption. 
There are few sources with apparent IR excess (BRRG 7, 89, 147, 148, 157, 152, 
169, 176), and few other that could be O stars with high reddening 
($A_{\rm V}>7$\,mag) (BRRG\,9, 13, 162). 
The absolute IR magnitudes and colours for OB ZAMS stars are adopted from 
Hanson, Howarth \& Conti (1997). 

The interpretation of IR CMDs is not straightforward as the early ZAMS stars 
locus is very steep, even steeper than that of the optical CMDs.
Moreover, the first two 
million years evolutionary tracks for massive stars are almost constant 
in colour.
Therefore, it is  extremely difficult to discern between a ZAMS and a 
main-sequence star using only near-IR photometric information. Thus, 
spectroscopic classification must be done, as 
an unreddened ZAMS O6.5\,V star looks like an unreddened O8\,V dwarf star 
using NIR broadband photometry. 

Parker et al. (1992) classified five stars in LH10 as ZAMS O stars
(O stars with \ion{He}{2} 4686 absorption line being stronger than any
other \ion{He}{2} absorption line).  According to these authors, the
stars PGMW\,3073, 3102, 3126, 3204 and 3264 are roughly coeval with but
slightly younger than the other typical ``non-ZAMS'' O stars.  Three of
these ZAMS stars are found in our survey, namely: PGMW\,3073 (O6.5\,V),
3102 (O7\,V) and 3126 (O6.5\,V). Only one star, PGMW\,3073 ($Ks=14.82$)
shows $Ks$ magnitude compatible with a 06\,V ZAMS star ($Ks=14.85$ for
a LMC distance modulus of 18.5). The other two stars (PGMW\,3102 and
PGMW\,3126) have $Ks$ magnitudes brighter than PGMW\,3073, in particular
PGMW\,3102 is almost one magnitude brighter in $Ks$. Their $Ks$ magnitude
would imply a ZAMS O3-O4 star.  According to the calibration of Vacca,
Garmany, and Shull (1996), the absolute magnitudes for O6.5\,V and O7\,V
stars stars are $-5.0$ and $-4.9$, respectively. So, the optical data
for these three stars ($M_V=-4.3, -5.4, -5.0$, for PGMW\,3073, 3102,
and 3126, respectively) would suggest that the $M_V$, and the spectral
type for PGMW\,3126 are in remarkable agreement, and PGMW\,3073 is 0.7
mag underluminous, and PGMW\,3102 is 0.5 mag overluminous. The model
for ZAMS stars predicts that they could be underluminous so that could
explain PGMW\,3073. It is possible to invoke different explanations to
take into account the departure in magnitudes for those ZAMS stars,
but in any case, all of this may be over-interpretation of the data;
in Figure 6 in Vacca et al. (1996), it is clear that the variation of
$M_V$ for a given spectral subtype can be quite large (they state a
$RMS=0.67$ for the deviation of the data ponts to the best fit in their
calibration). Therefore, the magnitude differences may be intrinsic
scatter in the absolute magnitudes. Also, we need to keep in mind the
intrinsic uncertainty in the spectral classification derived from spectra
with moderate signal-to-noise ratio and nebular contamination.

Another issue in the interpretation of NIR photometric information
is the effect of differential and internal reddening. Sources BRRG\,9, 13, and 
162 have the colours and magnitudes expected for early-O stars with high 
reddening. Different alternative explanations for these kind of sources can be 
suggested. They might be LMC red giants with moderate reddening but in a 
region with a peculiar extinction law, or Young Stellar Objects (YSOs) with 
intrinsic IR emission whose combined 
colours place them on the locus expected for stars with high reddening. 
We believe that the second alternative is the most plausible one. 
Rubio et al. (1998) reported similar IR sources in 30 Dor, and they discussed 
the importance of obtaining IR photometry in additional bands, 
such as $L$ and $M$, as well as high spatial resolution images and IR 
spectroscopy.

A direct comparison between the  NIR CMDs obtained for N11B
and the NE nebular filament in 30 Doradus (Rubio et al. 1998) 
indicates that sources with the strong IR excess found in the latter
are absent in the N11B region. In spite of a factor $0.5$ smaller 
area surveyed by Rubio et al. (1998) in that filament ($75''\times 75''$), 
they found six IR sources with $K_s<15$ and $H-K_s>1.0$. Our NIR survey  
centered in the LH10 association does not show any IR source 
with characteristics like those found in 30 Dor.
The most extreme IR sources discovered in this
survey, BRRG\,147 and BRRG 157, have $K_s>15$, and both are found 
towards the nebulosity where the CO cloud is located and where a methanol maser
has been reported.

Brandner et al. (2001), using HST/NICMOS images of the 30 Dor Nebula, 
suggested that sources with intrinsic IR excess $J-K_s>1$ and $J$ 
magnitudes, $17<J<19$, are intermediate mass Herbig Ae/Be candidates. 
Table ~\ref{candidates} lists the IR sources
in N11B that meet this criterion. It also lists the sources with $H-Ks>0.5$, 
including stars with interesting morphological association with the 
surrounding gas and dust, such as BRRG\,26, an embedded source in a compact 
nebulosity.
Candidates of classical T Tauri stars would have  $J \sim 20$
and $K_s \sim 18$ (Brandner et al. 2001) and are therefore  
beyond the sensitivity limit reached in our survey.

\subsection{Comparison between molecular CO and optical nebular emissions}

\begin{figure}
\figurenum{5}
\epsscale{1.00}
\includegraphics[angle=0,scale=0.9]{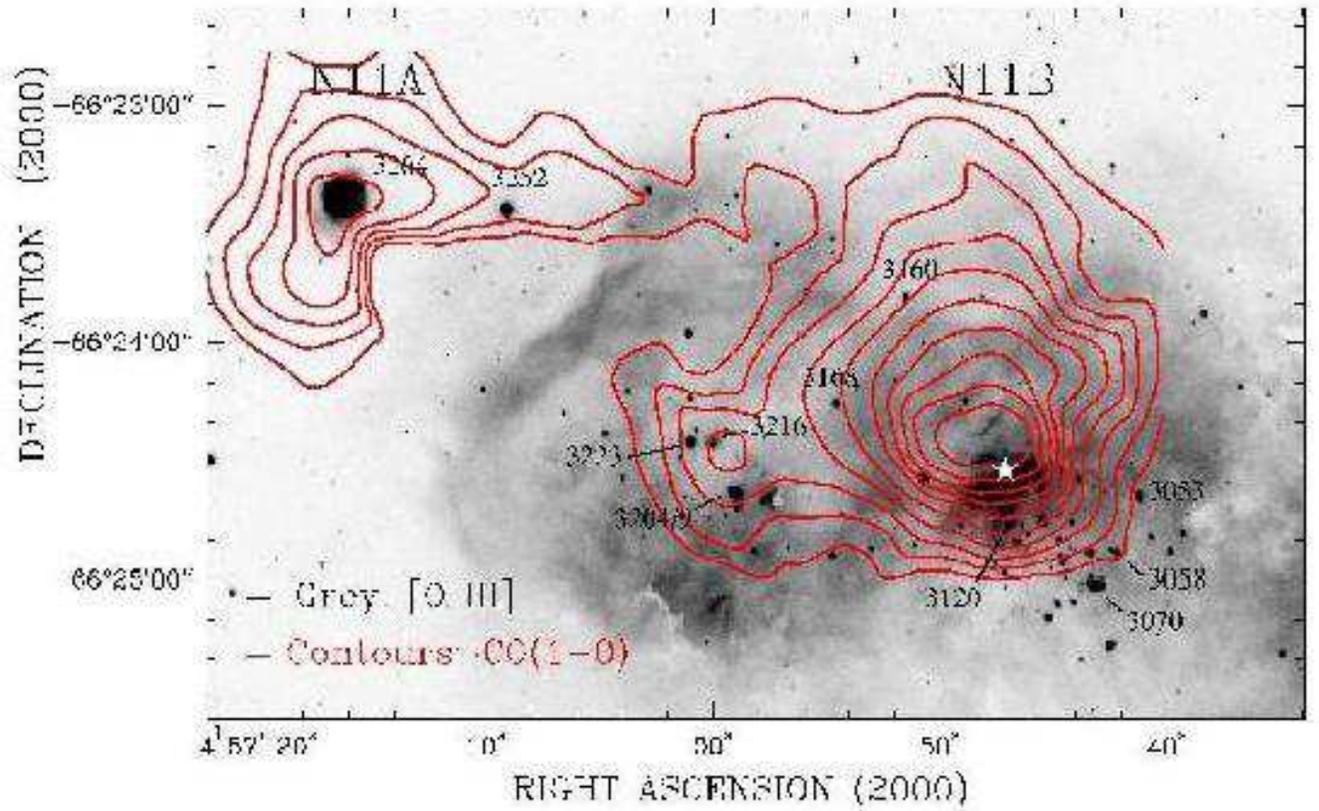}
\caption{Superposition of CO $(1\rightarrow0)$ contours on an NTT/EMMI 
[\ion{O}{3}] 
5007\AA\ image of N11A and N11B. Numbers denote optical stars from
Parker et al. (1992). The white star symbol shows the position of a methanol 
maser.
\label{fig5-co+o3}}
\end{figure}

Figure~\ref{fig5-co+o3} shows the distribution of the integrated 
CO$(1\rightarrow 0)$ emission superimposed over the 
[\ion{O}{3}] 5007\AA\ NTT image of the N11A and N11B region. 
The molecular gas is concentrated towards the brightest optical nebular 
emission features of these regions. Caldwell \& Kutner (1996) found that the 
CO molecular clouds in all of the N11 nebula was correlated to the H$\alpha$ 
emission.

In the particular case of N11B, the CO observations show that this emission 
is almost coincident in all its extension with the optical nebula mapped by
the high-excitation [\ion{O}{3}] gas. 
Therefore, most of the diffuse optical emission  could be a consequence
of photoevaporation and/or ionisation of gas at the interface of the
molecular cloud produced by the intense UV field generated by the 
LH10 association. 
There are several noticeable structures that suggest this scenario.
The peak of the CO emission in N11B is located to the north of the 
brightest optical nebular emission filament, and the ionising sources of 
LH10. 
This optical filament may be produced by the interaction of 
the O stars in LH10 with the border of the giant molecular 
cloud being eroded, resulting in a prominent PDR also seen in 
WFPC2 images (next Section).
This morphology clearly resembles the W and NE filaments in 30 Doradus 
(Barb\'a, Rubio \& Walborn 1999). 
Towards the molecular cloud, we find the reddest IR sources of our survey
and also a methanol maser reported by Ellingsen et al. (1994). Thus, this area 
is a prime candidate site for current star formation in N11B. 

A secondary CO emission peak is found to the east of N11B, in the direction 
of PGMW\,3216. Close to it, we find an O8.5\,IV star (PGMW\,3223) and a 
multiple system whose main component is an O3\,III(f*) star (PGMW\,3209),
reported by Walborn et al. (1999b). This molecular peak could be associated 
with the dusty cometary structures described by Naz\'e et al. (2001).

To the northeast of N11B, we find a separate nebular entity:
the N11A nebular knot 
(upper left quadrant in Figure~\ref{fig5-co+o3}). This compact \ion{H}{2} 
region has been studied by Heydari-Malayeri \& Testor (1985), and Parker et 
al. (1992) who proposed that an early O star (PGMW\,3264, O3-6\,V) is emerging 
from its protostellar cocoon. Heydari-Malayeri et al. (2001), using HST 
observations, found that this object is  a small compact group of five stars 
packed in a $2''\times2''$ area, dominated by a source of $y=14.69$, almost 
two magnitudes brighter than the second source in the group. 

CO emission is found towards the N11A nebular knot. The maximum of the CO 
emission is almost coincident with the position of
the compact group of stars, but shifted $10''$ to the south-east. 
Heydari-Malayeri et al. (2001) found a sharp nebular ridge to the north-east 
of the exciting stars indicating the presence of a PDR in the knot, 
favouring the interpretation of an interaction between the stars inside the 
optical knot and the molecular material mapped by the CO emission. 
The PDR lies on the edge of the molecular region facing the exciting stars, 
similar to the  case discussed previously for N11B. 

It is most interesting to study in detail the PDRs in N11A and N11B because
they might be in a different stage of evolution. In the case of N11A, the 
optical compact group of stars is visible indicating that the molecular cloud 
lies behind the stars and that they have evacuated a cavity towards our line 
of sight. This conclusion is supported by the fact that the O star has 
a relatively low reddening, $A_V=0.6$, probably due only to dust mixed with
the ionised gas (Heydari-Malayeri et al. 2001). 
These authors proposed that the nebular 
morphology of N11A is a good example of the champagne model (Tenorio-Tagle 
1979; Bodenheimer et al. 1979), in the stage when the newborn stars disrupt
the molecular cavity. 
Small nebular emission filaments discovered
by Heydari-Malayeri et al. (2001) are located on the south-west border of
the nebular knot, and show arcs pointing toward the direction where
the CO emission has a sharp edge, indicating an abrupt drop of the 
molecular gas density. This morphology suggests that the compact group 
of stars could be blowing this side of its stellar nursery, in a similar 
morphological picture to that seen towards Knot 1 in the 30 Dor Nebula 
(Rubio et al. 1998; Walborn et al. 1999a; Walborn, Ma\'{\i}z-Apell\'aniz \&
Barb\'a 2002).

\begin{figure}
\figurenum{6}
\epsscale{0.5}
\plotone{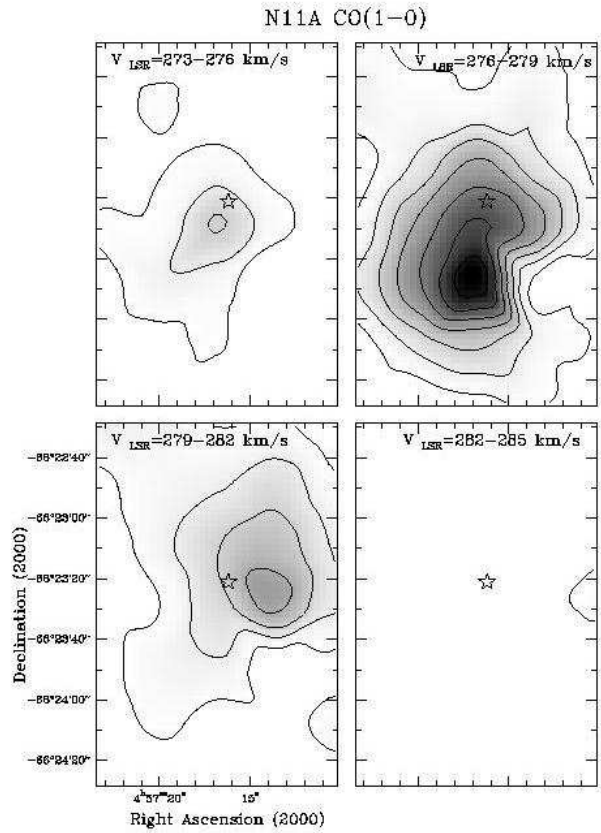}
\caption{Integrated CO velocity channel maps of N11A. The velocity range in
each channel is 3 \kms\ starting at $V_{\rm LSR}=273$ \kms. Contour levels 
go from 0.5 K\kms\ to 3.7 K\kms\ by 0.5 K\kms. The position of PGMW\,3264, 
is indicated by a star in each channel map.
\label{fig6-n11a-co}}
\end{figure}

\begin{figure}

\figurenum{7}
\epsscale{0.5}
\plotone{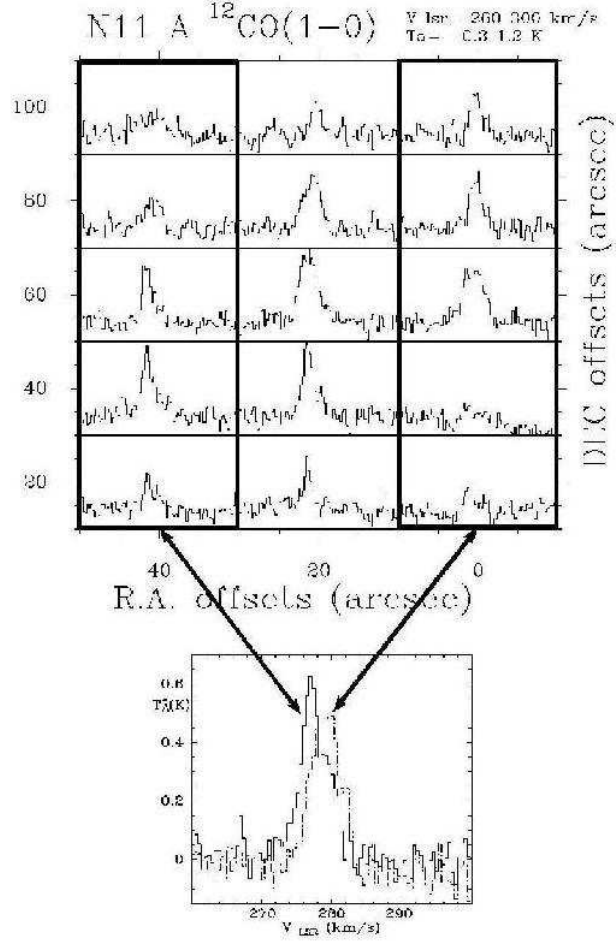}
\caption{CO spectra of the N11A region. The molecular gas
shows a different CO velocity in the eastern side of PGMW 3264
with respect to its western side. This is better seen in the lower
panel where composite spectra of the eastern side and western side are
shown. The western side composite spectrum is drawn in dotted lines
while the eastern composite spectrum is drawn in continuous line. The velocity
difference is 2 \kms. The temperature scale of the spectra are in antenna 
temperature $T^*_A$, and thus should be multiplied by 1.4 to obtain them 
in $T^*_R$.
\label{fig7-n11a-co-sp}}
\end{figure}

The velocity information of the CO gas supports this scenario. 
Figure~\ref{fig6-n11a-co} shows four different velocity integrated channel 
maps each with 3 \kms\ range. The position of the compact group of massive 
stars is 
indicated by a star in each panel. The CO emission peaks to the south of the 
stars in the velocity range $V_{\rm{LSR}}=276-279$ \kms\ while it peaks to 
the west at the velocity range $V_{\rm{LSR}}=279-282$ \kms. So, there is gas 
spatially concentrated at different velocities. Figure~\ref{fig7-n11a-co-sp} 
shows the CO spectra around the compact cluster. 
We have integrated the 
emission to the east and west of the cluster, as indicated in the figure, 
and the resultant CO spectra clearly show a velocity separation of 2 \kms. 
Thus, the molecular gas is being accelerated and we see either an expanding 
envelope or the interaction of the cluster and the molecular gas which has 
produced a cavity. This cavity is formed in the thinner region of the parental 
molecular cloud. 
The molecular gas towards this direction has a greater velocity than that 
towards the densest part of the molecular cloud. We plan to obtain more 
sensitive and higher spatial resolution CO$(2\rightarrow 1)$ observations 
of this region. 

The CO emission distribution between N11A and N11B is devoid of molecular
gas (Figure~\ref{fig5-co+o3}). 
There is a sharp edge in the molecular cloud west of N11A
and then a huge elongated area of about $30\times10$\,pc where
no molecular emission has been detected. 
This molecular gas void is coincident
with the one present in the optical nebular emission. The nebular emission 
to the east of N11B shows a smooth surface as seen in the [\ion{O}{3}] gas, 
suggesting that an energetic event could  have taken place producing the 
cavity, pressing the molecular gas towards N11A and mechanically 
smoothing, by some shock interaction, the eastern wall of the molecular 
cloud N11B.
The compression of the molecular cloud in N11A could have favoured the
formation of a dense core that gave origin to the compact star cluster
(PGMW\,3264). Alternatively, the energetic event could have favoured 
the gas evacuation around the new born stars.

Sensitive and better spatially sampled observations of the molecular gas
in N11B using the CO$(2\rightarrow1)$ emission line are underway. 
Preliminary results show that the distribution of the molecular gas is 
similar to the one found in CO$(1\rightarrow0$). There seems to be a different 
CO$(2\rightarrow1)$/CO$(1\rightarrow0)$ ratio towards the peak of N11B
(Rubio M., private communication).

\subsection{A Methanol Maser}
\label{methanol}

Methanol (CH$_3$OH) maser emission at 6.67 GHz is one of the strongest
astrophysical masers, and therefore it has been possible to detect it
in extragalactic environments. This maser emission has been established as 
a tracer of star-forming regions often associated to ultra-compact \ion{H}{2}
regions in our Galaxy (e.g. Caswell et al. 1995). Three methanol masers 
were discovered in the LMC, all of them placed in \ion{H}{2} regions, 
namely: N105a (Sinclair et al. 1992), N11B (Ellingsen et al. 1994), and
DEM52 (Beasley et al. 1996). The methanol maser associated with 
the N11B nebula is the second one in intensity, with a peak flux density
of 0.3 Jy, and a heliocentric velocity of 301 \kms ($V_{\rm LSR}=287$). 
The coordinates reported by Ellingsen et al. (1994) for this maser are 
$\alpha(2000)=4^{\rm h}56^{\rm m}47^{\rm s}12$, 
$\delta(2000)=-66^\circ 24'31\farcs8$, 
and it was  unresolved with a $2\farcs5$ beam. 

The N11B maser, according to the  given radio position, is located 
in one of the dusty prominences where our NIR images show 3 stellar IR sources
inside the $2\farcs5$ ATCA beam (BRRG\,144, 147, and 148).
Figure~\ref{fig2-ir-n11b-zoom} shows the maser position marked by a circle.
Only one of these IR sources, BRRG\,144, has been identified
with an optical stellar source, PGMW\,3123. This star has been classified as 
a late-O type star, O8.5\,V, with usual stellar IR properties.
Its IR magnitudes are $Ks=14.66$, $J-H=0.03$, $H-Ks=0.08$  with
an expected reddening of $E_{B-V}=0.2$. This O star probably
belongs to the LH10 association and is probably not associated with the maser.
The two  other IR stellar-like sources are among the reddest sources in the 
field, (BRRG\,147, $Ks=15.48$,  $J-H=0.86$,  $H-K=1.22$; 
        BRRG\,148, $Ks=16.72$,  $J-H>0.88$,  $H-K=1.20$). 
They show an intrisic IR excess and meet the Brandner et al. (2001) criterium 
for intermediate mass Herbig Ae/Be star candidates (Section~\ref{photom_ir}). 
In the Milky Way, Walsh et al. (1999) did a NIR survey 
towards a selected sample of methanol maser and/or ultracompact \ion{H}{2}
regions, and they found that about 50\% of the methanol masers 
have associated NIR counterparts. The IR counterparts were identified not 
only from coincidence in position, but also from the fact that 
they show the reddest colours in the sample. 
A similar scenario is found between the association of the IR sources BRRG\,147
and 148 with the methanol maser in N11B. 

\subsection{Comparison with narrowband WFPC2 images}
\label{wfpc2}

Aiming to detect optically wind-blown bubbles in the interstellar 
medium of N11B, Naz\'e et al. (2001) obtained a set of WFPC2 images in F502N 
([\ion{O}{3}]) and F656N (H$\alpha$) narrowband filters of such nebula. 
These images are
a superb set to compare morphological aspects between the ionised 
features, molecular clouds and infrared sources distribution. 

A false colour image (Figure~\ref{fig8-n11-wfpc2}) of N11B was produced 
by combining  H$\alpha$ (F656N) and [\ion{O}{3}] (F502N) WFPC2 
mosaics.  H$\alpha$  was mapped in red and [\ion{O}{3}] in blue, 
the green channel corresponds to H$\alpha$ and [\ion{O}{3}] together.
In this image, nebular features with high H$\alpha$ to 
[\ion{O}{3}] ratio stand out in yellowish colours, while those  
brighter in [\ion{O}{3}] appear blue.
Figure~\ref{fig8-n11-wfpc2} shows that strongest the nebular emission 
comes from the PDR interface between the molecular gas and the ionised 
cavity, where several LH10 stars are located. ie. PGMW\,3128, 
PGMW\,3120. This PDR is in fact a dusty prominent feature that shows a rugged 
surface with many bright and dark protuberances, some of them 
facing to the hot stars. It also seems to be illuminated 
from behind due to the presence of a weaker PDR edge seen to the north east.
The PDR emission pattern looks like the heads of pillars in M16 
(Hester et al. 1996) as it was described by Naz\'e et al.(2001). 
These dusty features coincide with the the maximum of the CO emission 
as seen from the CO distribution in Figure~\ref{fig5-co+o3}.

Figure~\ref{fig9-n11-pilar} shows a close-up of the dusty prominence and its
neighbouring stars of the N11B WFPC2 
colour mosaic display in Figure~\ref{fig8-n11-wfpc2}.
The reddest IR sources (black numbers in Figure~\ref{fig9-n11-pilar}) 
detected in our images are lying toward the molecular cloud core,
as expected for a star forming region signature.
Some of the IR sources are  associated with small dusty protuberances, 
such as BRRG\,108, 113 and 129, while others are close to the bright H$\alpha$ 
spot (sources BRRG\,162 and BRRG\,165). The methanol maser 
(marked with a blue circle) is located in a dark spot close to the weak PDR 
edge north of the main molecular protuberance, between sources BRRG\,147 and 
BRRG\,148. 

On the surface of the molecular cloud, dark finger tip-like features are
seen with bright rimmed borders and dark tails (northwest corner of 
Figure~\ref{fig9-n11-pilar}). 
The scale of these features is about of $1'' \times 0\farcs5$ 
($0.25 \times 0.12$ pc). Two of them have IR sources, BRRG\,108 and
BRRG\,129, directly associated. 
Their appearance is similar to those discovered 
by Scowen et al. (1998) (see also Barb\'a et al. 1999) in the inner cavity 
of the 30 Dor Nebula. Those in N11B possibly present a different illumination 
pattern.
Walborn et al. (1999a) discovered IR sources embedded 
in dark globules in the 30 Dor region. Thus, N11B would be the second
region in the LMC where these finger-like features with IR sources 
are detected.
The two sources associated with these features, BRRG\,108, and 129, 
plus BRRG\,113 (very close to BRRG\,108) satisfy the criterion suggested 
by Brandner et al. (2001) for intermediate-mass Herbig Ae/Be candidates 
in 30 Dor.

To the south of the PDR, there is a well defined nebular ring seen in bluish
colour in Figure ~\ref{fig9-n11-pilar} around PGMW\,3120. 
It consists of a segmented 
nebular arc of about $3''-6''$ radius.  
Although, Naz\'e et al. (2001) did a detailed nebular kinematic study of 
N11B, defining  several shells from the H$\alpha$ and
[\ion{N}{2}] emission line splitting, they found that this nebular ring was 
not detected as a kinematical feature in their echellograms. Their kinematic 
analysis around PGMW\,3120 reveals a huge shell 
extending to the south, east, and west as seen in the splitting of the 
nebular line emission.
Towards the north of PGMW\,3120, Naz\'e et al. (2001) detect the emission 
line splitting, but at a distance of $7''$, where the bright ring is located, 
the lines do not show any splitting further north (see their Figure~10).
They suggest that the small H$\alpha$ arc is probably  ``fortuitous''.

We propose an alternative interpretation for this nebular arc.
Photodissociation regions are characterised not only by the presence 
of a molecular cloud surface that is being photodissociated and photoionised,
but also by the existence of a photoevaporative flow. This flow has been
detected in several PDRs where the strong radiation field is
acting on the molecular cloud surface, and a clear example of this
phenomenon are M16's Elephant Trunks (Hester et al. 1996). As we have 
mentioned earlier, the HST/WFPC2  images of N11B show the same morphological 
pattern in the nebular emission close to the dusty prominence.  
The H$\alpha$ and [\ion{O}{3}] images show clear striation patterns 
peperdicular to the molecular interface, where  photoevaporated gas 
is streaming away from 
the molecular cloud toward the ionised \ion{H}{2} region. 
This striation pattern 
is clearly seen in N11B, despite of the factor 25 greater in distance between
M16 and the LMC. Thus, the ring around PGMW\,3120 is probably due to the 
interaction between the stellar winds and the photoevaporative outflow which 
produces a concentration of gas at the region where the winds and the 
photoevaporation flow interact. This would be the case in the northern region 
of PGMW\,3120. At the position of the optical nebular arc there is no cavity, 
and the nebular spectrum does not show further splitting. To the north of the 
ring the region is dominated by the photoevaporative flow and not by 
the stellar winds that are responsible for the splitting seen in the nebular 
lines in all other directions. The ``kinematic expanding shell'' proposed by 
Naz\'e et al. (2001) does not continue to the north of PGMW\,3120.

\subsection{PDR and other nebular structures}

The structure of the PDR's in H$\alpha$ and [\ion{O}{3}] emission 
lines accross PGMW\,3120 is mapped along an almost north-south cut 
in Figure~\ref{fig10-pdr-profile}. The emission
line surface brightness (in detector counts) shows distinctive features.
Three main PDRs are identified across the line and labelled
with $I1$, $I2$, and $I3$. 

The PDR $I1$ shows a typical ionisation structure
where H$\alpha$ emission appears more deeply concentrated in the
interface than the [\ion{O}{3}] emission 
line which peaks at the outer edge of the interface. 
Unfortunately, there are no [\ion{S}{2}] HST images yet, but we would
expect filamentary structures in such images, similar to 
those seen in the 30 Doradus region (Scowen et al. 1998; Rubio et al. 1998).

The PDR interfaces $I2$ and $I3$ are facing each other indicating 
the presence of a ionised cavity in between. This cavity 
is also detectable in the kinematic profile
presented by Naz\'e et al. (2001). In their Figure~10 it is possible to see
 a strong double-peak emission line profile developing at  $10''$
accross the region (see PGMW\,3223 EW profile). 
The star PGMW\,3123 (O8.5\,V) is
located in the cavity and is perhaps contributing with UV photons. 
There may be
some other hot stars hidden in the dusty prominence which appears similar
to a wide pillar feature.

Following the  emission line spatial structure of the PDR from the 
interface $I1$ to the south,
the arc around PGMW\,3120 (labelled $A1$) is found, as well as 
a deep depression in both the H$\alpha$ and 
[\ion{O}{3}] flux distribution immediately south of the stars. 
This spatial distribution 
suggests that the gas density around the stars is depleted to the south.
To the north, the stellar winds are interacting with the PDR outflow. 
This region could be an interesting area in which to investigate the energy 
balance and 
deposition energy of early O stars in a pristine \ion{H}{2} region with 
well developed PDR interfaces.

\begin{figure}
\figurenum{8}
\epsscale{1.00}
\plotone{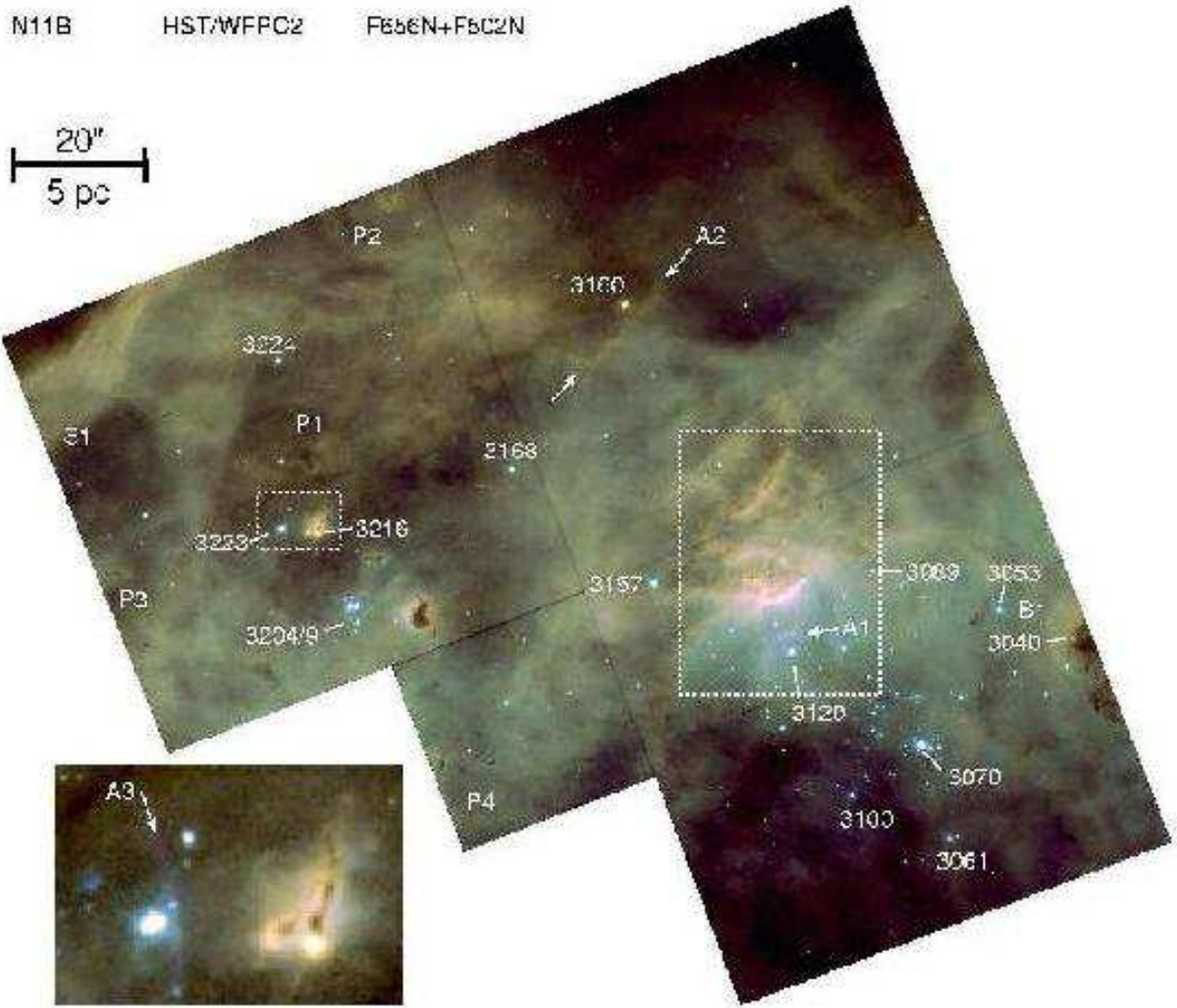}
\caption{WFPC2 composite color image of N11B. The F656N (H$\alpha$) and 
F502N ([\ion{O}{3}]) images
are in the red and blue channels, respectively, while the green channel 
is a combination of both filters. Numbers denote the optical stars of 
Parker et al. 1992. Labels $A1$, $A2$ and $A3$ mark the position of nebular
arcs discussed in the paper. Label $B1$ indicates a Str\"omgren-like sphere
around PGMW\,3040. Labels $P1$ to $P4$ show finger-like features pointing
to the PGMW\,3204/09 stellar group. The dashed box to the right around the
pillar is displayed in detail in Fig.~9. The dashed box to the left including
PGMW\,3223 and 3216 is zoomed in the lower panel.
\label{fig8-n11-wfpc2}}
\end{figure}

\begin{figure}
\figurenum{9}
\epsscale{0.85}
\plotone{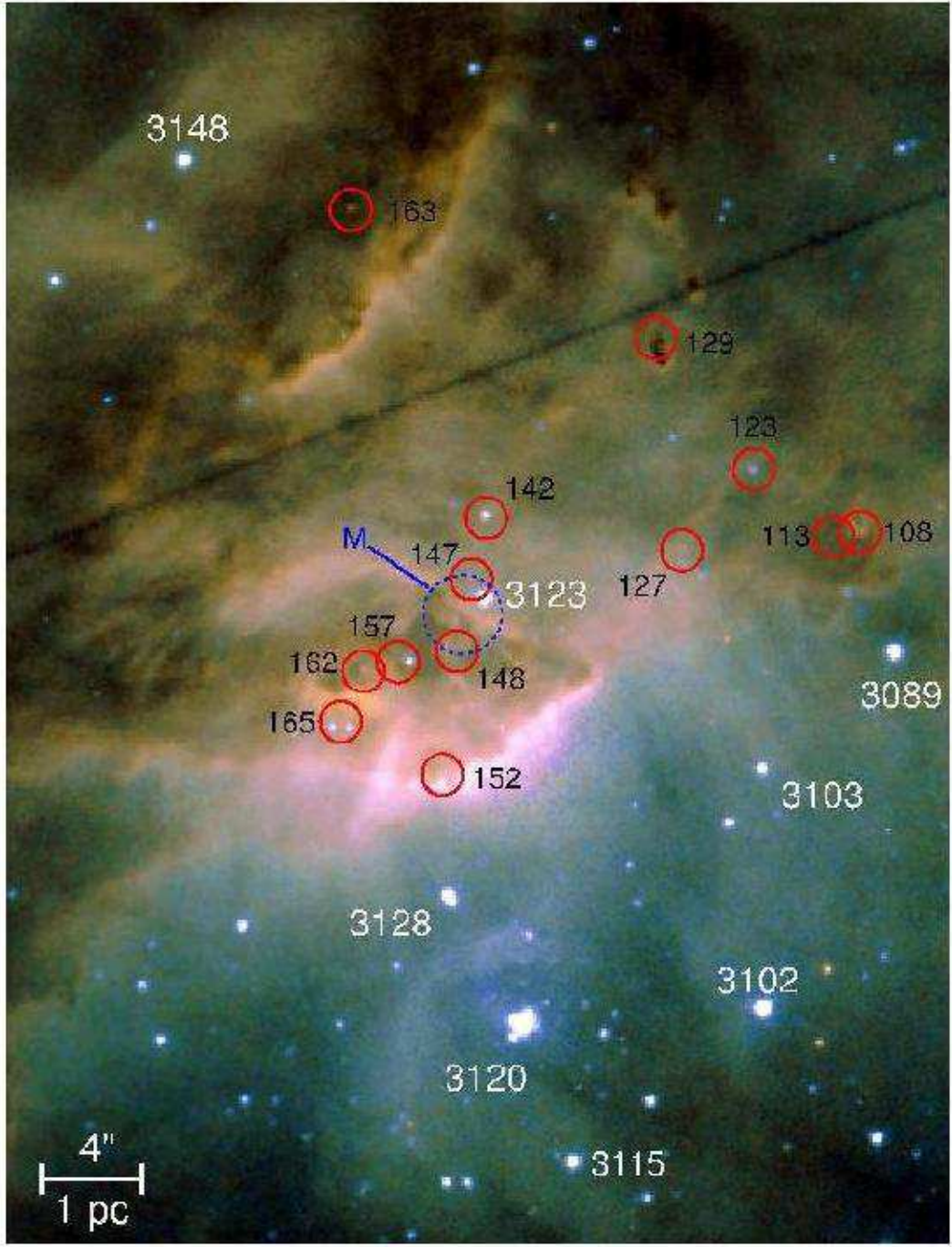}
\caption{Detail of the WFPC2 composite color image (Fig. 8) around the
dusty prominence. Numbers have the same meaning that in Fig.~1. The M
label denotes the position of the methanol maser.
\label{fig9-n11-pilar}}
\end{figure}

\begin{figure}
\figurenum{10}
\epsscale{1}
\includegraphics[angle=-90,scale=0.7]{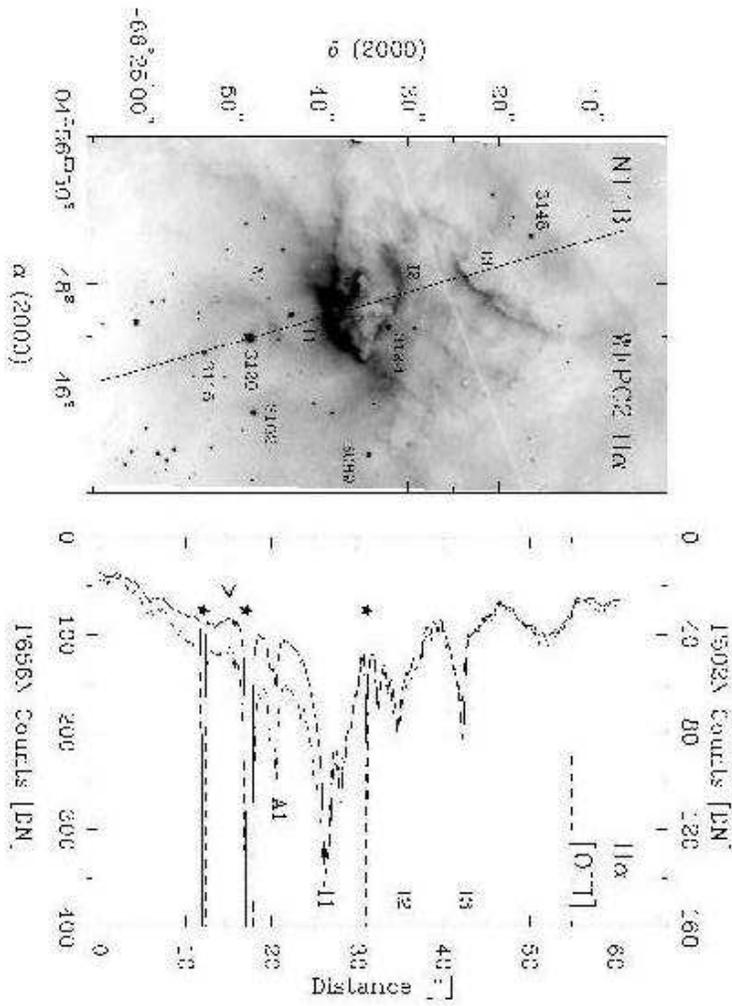}
\caption{WFPC2/F656N image in the region of the dusty pillar in N11B. 
The $60''$ line crossing the image from the south to the north indicates the
spatial profile plotted in the left panel. Labels $I1$, $I2$ and $I3$
mark prominent PDRs. Label $A1$ is the nebular arc around PGMW\,3120.
The arrow indicates the position of the depression in the H$\alpha$ and 
[\ion{O}{3}] emission immediately to the south of PGMW\,3120. Strong 
spikes in the plot are stars (labelled with black stars).   
\label{fig10-pdr-profile}}
\end{figure}

Other interesting features in the N11B region are
two arcs labelled as $A2$ and $A3$ in Figure~\ref{fig8-n11-wfpc2}. 
Naz\'e et al. (2001) suggested  that arc $A2$ could be identified as 
part of a ring nebula around the K\,I star PGMW\,3160, and they proposed 
that the optical filaments were not associated with dusty features. 
We believe that such filament could well be another PDR facing the 
main group of hot stars of the LH10 association. This PDR appears 
to run parallel to the $I3$ PDR interface described above. 
In a close inspection of the WFPC2 images, the $A2$ arc shows the same spatial 
structure as the
 $I3$ PDR but with a fainter emission. In Figure~\ref{fig8-n11-wfpc2} 
the $A2$ arc stands out as a yellowish colour filament indicating a higher 
H$\alpha$ to  [\ion{O}{3}] emission ratio (as also noted by Naz\'e et al.
2001). The distribution of PDRs in this part of the molecular
cloud looks like the bright rimmed borders of a thunderstorm cloud
illuminated at the sunset. The dark region of the image to the north 
of PGMW\,3160 seems to be in the shadow of the molecular cloud, protected from 
the UV light produced by the hot stars of LH10.

A tiny arc, labelled $A3$ in the inset of
Figure~\ref{fig8-n11-wfpc2},  appears just close to 
PGMW\,3223\footnote{This star is resolved in two components in WFPC2 
images, with an intensity ratio in H$\alpha$ filter of about 1:0.6 
and a separation of $0\farcs3$)} an O8.5\,V star. 
The arc is located in between the
stars and the Y-shaped cometary dust clouds described by Naz\'e et al.
(2001). This arc could be of similar nature as that seen around 
of PGMW\,3120. The outflow pattern seen around the east branch of
the Y-shaped cloud suggests that PGMW\,3223 could contribute to its
photoevaporation. Naz\'e et al. (2001) identified a star with strong
H$\alpha$ emission in the tip of the Y-shaped dust cloud and
suggested that this star is part of the ongoing star formation
in the region. The star, identified as PGMW\,3216, has  
$V=15.52$, $B-V=+0.40$, $U-B=-0.51$, and $Q=-0.80$ indicating that it is
an obscured hot source. A secondary maximum of the CO emission
is located in this region (see Figure~\ref{fig5-co+o3}), favouring the
scenario of star-forming activity in such place.

\begin{figure}
\figurenum{11}
\epsscale{1}
\includegraphics[angle=0,scale=0.7]{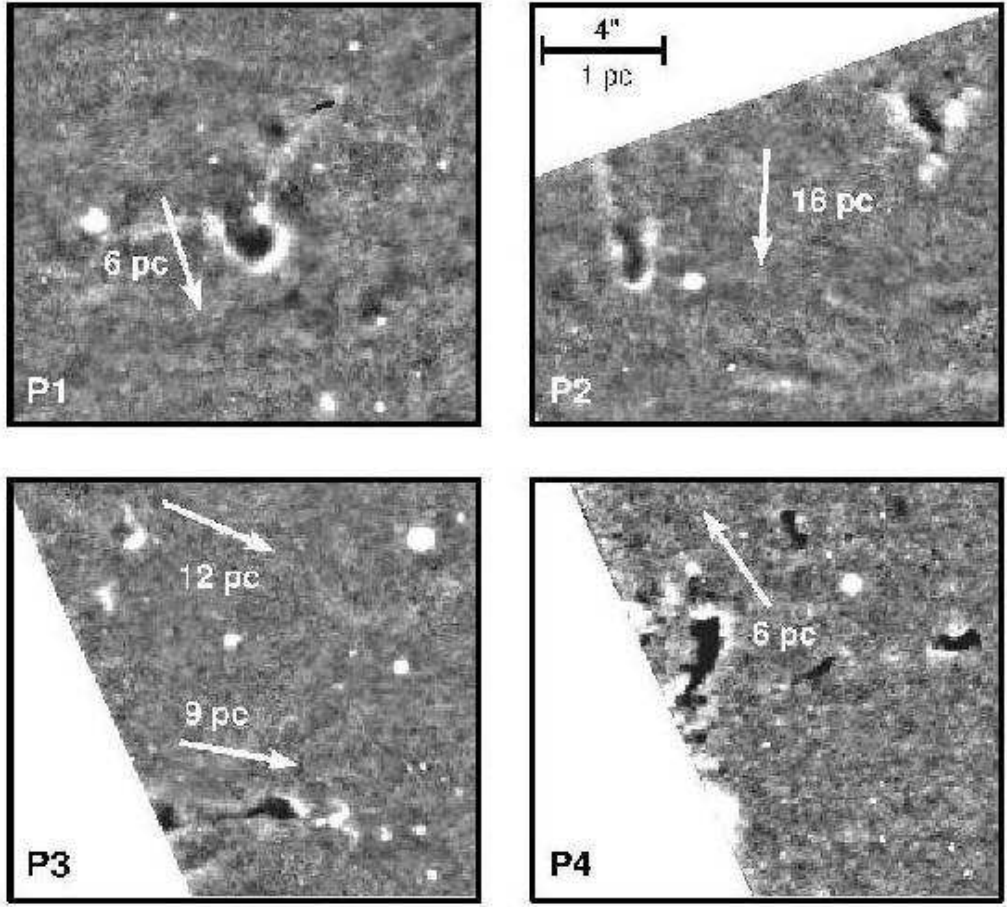}
\caption{The rimmed finger-like features $P1$, $P2$, $P3$ and $P4$.
This Figure was done by subtracting a 15 pixels median image from the 
WFPC2/F656N image, and allows to enhance the contrast 
between the bright and dark areas, eliminating the diffuse emission component. 
The arrows point to the direction to PGMW\,3204/09 star group for each 
feature, 
and the numbers indicate the projected linear distance to those stars.
\label{fig11-n11-fingers}}
\end{figure}

Pointing toward the stars PGMW\,3204/09 there are several
dusty clouds, some of them showing PDR interfaces. The kiwi shaped
cloud at $10''$ to the west of those stars was nicely described 
by Naz\'e et al. (2001). The rimmed finger-like features with 
bright tips, labelled  $P1$, $P2$, $P3$ and $P4$ in 
Fig.~\ref{fig8-n11-wfpc2} are shown in Fig.~\ref{fig11-n11-fingers}.
They show an illuminated surface facing PGMW\,3204/09  
possibly due to the UV radiation engraving the molecular cloud 
as far as $60''$ (15pc) from the stars. This Figure was done 
by subtracting a 15 pixels median image from the F656N image. This
procedure allows us to enhance the contrast between the bright and
dark areas, eliminating the diffuse emission component. The arrows point
to the direction to PGMW\,3204/09 star group for each feature. It also
indicates the projected linear distance to those stars.

There are dusty filamentary structures which don't seem to be directly 
illuminated by the hot stars, located to the south and east of PGMW\,3070 
(see Fig.~\ref{fig8-n11-wfpc2}). The presence of these dusty clouds could be
relevant for the cluster studies as they can be responsible for spacial 
variable extinction at the scale of few tenths of arceseconds in the LH10 area.

A very filamentary cloud appears close to the the west corner of the WFPC2 
images with an embedded star labelled 
$B1$ in Fig.~\ref{fig8-n11-wfpc2}. This is the BRRG\,26 source ($Ks=15.88$, 
$J-H=0.46$, $H-Ks=0.19$) in our IR images which shows a compact nebulosity. 
The source is embedded in the tip of the dark pillar which 
appears as a yellowish (H$\alpha$ emission) semicircle of
about $5''$ (1.25 pc) radii,  producing a Str\"omgren-like sphere.
It is also identified as PGMW\,3040 ($V=17.69$).

In the east corner of Fig.~\ref{fig8-n11-wfpc2} (labelled with $S1$) 
there are several parallel filaments. These filaments could be the surface
of the molecular cloud facing the cavity described in Section~3.3.

\subsection{Other early O stars in LH10}

The LH10 association is very rich in O stars and it is obvious that 
the WPFC2 images help to resolve tight stellar systems in some objects, like
PGMW\,3204/09 (Walborn et al. 1999b) and PGMW\,3264 (N11A, 
Heydari-Malayeri et al. 2001) and PGMW\,3120, PGMW\,3223 in this paper.

PGMW\,3120 is a very interesting object. The star was classified by 
Parker et al. (1992) as O5.5\,V((f*)), with the peculiarity of 
displaying \ion{N}{4} 4058\AA\ emission and \ion{N}{5} 4604\AA\ absorption 
simultaneously, two lines which are not expected to be present in an O5.5\,V  
spectral type star. In the WFPC2 narrowband images, PGMW\,3120 appears 
composed of 3 stars of similar brightness (1:0.88:0.82
in H$\alpha$) forming almost an equilateral triangle with $0\farcs35$ 
of separation between them. The peculiar spectral classification 
could be explained if one of the brightest components is an O3 star
\footnote{Recently, a new spectrum of PGMW\,3120 obtained at ESO
by D. Lennon was analysed by N. Walborn, who couldn't detect \ion{N}{4}
emission and \ion{N}{5} absorption above the noise level, discharging
the possible presence of an O3 star. Additional spectroscopic observations 
are planned for this interesting object.}.

Heydari-Malayeri \& Testor (1983) indicated that PGMW\,3070 ($\alpha$ object 
in their paper) has probably a multiple nature. Parker et. al (1992) 
concluded that the star is resolved into 5-7 components with an integrated 
spectral type O6\,V. In our IR images 9 stars are found  in a circle of 
$4''$ (1\,pc) radii.  Their IR magnitudes and colours look like an OB open 
cluster with low reddening. Three of these are bright IR sources, 
BRRG\,75, 67, and 61, in order of brightness. The WFPC2 images show 18 
components in the same area where 9 are seen in the IR. PGMW\,3070 
is resolved in four components packed in a circle of $0\farcs5$ (0.12\,pc) 
radii. 

PGMW\,3061 was classified by Parker et al. (1992) as a classic example of 
O3\,III(f*) giant star with a possible faint blend (PGMW\,4020), 
with $\Delta V=3.08$. In our IR images the object appears as a double star, 
BRRG 146,  with a relatively bright IR close companion, BRRG\,49 at $1\farcs4$, 
and $\Delta Ks=0.92$. 

PGMW\,3126 (O6.5\,Vz) and PGMW\,3128 (O9.5\,IV) are both double stars 
with faint companions at about $0\farcs3$ in WFPC2 images. The IR 
counterparts of these O stars are seen as the single star BRRG\,146, and 
BRRG\,150, respectively. 

All these O stars show normal IR colours in our colour-magnitude 
and colour-colour diagrams.

\section{Discussion and summary}

We confirm that a second generation of stars is currently forming in the 
N11B region.
Our IR images show the presence of several bright IR sources which appear
located toward the molecular cloud as seen from the CO emission in the area.
Several of  these sources show IR colours with YSO characteristics and they are
prime candidates to be intermediate-mass Herbig Ae/Be stars following
the  criterion of Brandner et al. (2001). 
For the first time, an extragalactic methanol maser is directly associated 
with IR sources embedded in a molecular core. Two IR sources, BRRG\,147 and 
148, are found at $2''$ (0.5 pc) of the methanol maser reported position. 
They are among the reddest IR sources in N11B and therefore the strongest 
candidates of YSOs. IR spectroscopy of these sources will reveal their nature.

The overall picture of N11B including the CO and ionised gas distributions 
clearly demonstrate that the nebula is a region of interaction between 
stellar winds and UV radiation from hot young stars, as well as of PDRs 
in the molecular clouds originated by the action of such stars.
Almost all the optical nebular emission in N11B is directly associated with 
CO emission, suggesting that the main source of such optical emission 
comes from the molecular cloud surface, where the gas is photoevaporated 
and ionised. In fact, all the brighter optical nebular emission filaments
and peaks are PDRs facing the hot stars: a bright rim $10''$ to the north 
of the multiple O star PGMW\,3120, a Y-shaped cometary globule close 
to the double O star PGMW\,3223, a kiwi-shaped globule $10''$ to the west 
of the multiple early-O star PGMW\,3204/09

The nebular knot N11A to the east of N11B is also associated with a molecular
cloud core. This knot is excited by the multiple O star PGMW\,3264, and we find
evidence of the interaction of the stellar winds and the
molecular gas. This nebula has the morphology and
kinematics of a champagne flow where the massive stars have
open a cavity in their parental molecular cloud. Future IR data will allow 
to search for IR embedded sources in the region.

Walborn \& Parker (1992) proposed that N11 was few $10^5$ years older than 
the 30 Doradus, based on their analysis of the relative ages of the stellar
content of both giant \ion{H}{2} regions. 
In 30 Doradus, a new generation of massive stars is currently being 
formed in the peripherical 
molecular clouds apparently due to the energetic activity of the hot massive 
stellar core R136 (Rubio et al. 1998; Walborn et al. 1999a; Brandner et al. 
2001). 
The stellar content of such new generation in 30 Doradus has
O stars embedded in compact nebulosities (Walborn et al. 2002 and references 
therein), with a relative age of about 1-2 Myr younger than the core R136, 
characterised by the presence of luminous H-burning WN stars with an age 
of about 2 Myr. Walborn et al. (1999b) derived an age of about 3.5 Myr 
for the LH9 association in the N11 core, and less than 1 Myr for PGMW\,3209 in 
LH10. Their results confirm the youth of the LH10 association (also deduced 
from absence of WR stars), and that the N11 region is older than 30 Doradus.

Our study supports the idea that N11B is in a later stage of evolution
than 30 Doradus. It shows a lack of bright IR sources compared to the 
star forming regions in 30 Dor. Our brightest IR source has $K=15$, 
while several $K=12$ IR embedded stars were observed in 30 Dor.
The IR sources in N11B are probably intermediate mass star candidates
still embedded in the molecular gas. The O stars have blown away the 
molecular material and are disrupting the molecular cloud surface.
Thus, we may be witnessing the last stage of the second generation
burst in N11B.
 
\acknowledgements

M.R. is supported by the Chilean {\sl Center for Astrophysics} 
FONDAP No. 15010003. This work was funded by FONDECYT (CHILE) through 
grants No 1990881 and No 7990042.
R.H.B.  gratefully acknowledge support from the Chilean 
{\sl Center for Astrophysics} FONDAP No. 15010003 and from Fundaci\'on
Antorchas, Argentina (Project No 13783-5).
We wish to thanks to the referee Joel Parker 
for the corteous and useful comments
that greatly improved the presentation of this paper.
This publication makes use of data products from 2MASS, which is a 
joint project of UMass and IPAC/Caltech, funded by NASA and NSF.

\clearpage
\references

Arnal, E. M., Cersosimo, J. C., May, J., Bronfman, L. 1987, A\&A, 174, 78

Barb\'a, R. H., Walborn, N. R., \& Rubio, M. 1999, RMxAAC, 8, 161

Beasley, A. J., Ellingsen, S. P., Claussen, M. J., \& Wilcots, E. 1996, 
        ApJ, 459, 600

Bodenheimer, P., Tenorio-Tagle, G., \& Yorke, H. W. 1979, ApJ, 233, 85

Brandner, W., Grebel, E. K., Barb\'a, R. H., Walborn, N. R., \& Monetti, A. 
        2001, AJ, 122, 858

Caldwell, D. A., \& Kutner, M. L. 1996, ApJ, 472, 611

Caswell, J., Vaile, R., Ellingsen, S., Whiteoak, J., \& Norris, R. 1995, 
        MNRAS, 272, 96

Cutri, R.M. et al. 2000, Explanatory Supplement to the 2MASS Second
Incremental Data Release (Pasadena: Caltech)

Davies, R. D., Elliot, K. H., \& Meaburn, J. 1976, MNRAS, 81, 89

Ellingsen, S. P., Whiteoak, J. B., Norris, R. P., Caswell, J. L., \&
        Vaile, R. A. 1994, MNRAS, 269, 1019

Elmegreen, B. G. 1998, in ``Origins'', ASP Conf. Ser. 148, C. E. Woodward, J. 
        M. Shull, \& H. A. Thronson, Jr. (Provo: ASP), 150   

Elmegreen, B. G. \& Lada, C.J. 1977, ApJ, 214, 725

Hanson, M. M., Howarth, I. D., \& Conti, P. S. 1997, ApJ, 489, 698

Henize, K. G. 1956, ApJS, 2, 315

Hester, J. J.  et al. 1996, AJ, 111, 2349
 
Heydari-Malayeri, M., \& Testor, G. 1983, A\&A, 118, 116

Heydari-Malayeri, M., \& Testor, G. 1985, A\&A, 144, 98

Heydari-Malayeri, M.,  Royer, P., Rauw, G., \& Walborn, N. R. 2000, A\&A, 361, 877

Heydari-Malayeri, M., Charmandaris, V., Deharveng, L., Rosa, M. R., Schaerer,
        D., \& Zinnecker, H. 2001, A\&A, 372, 527

Israel, F. P. \& de Graauw, Th. 1991, in ``The Magellanic Clouds'', 
        IAU Symposium No. 148, R. Haynes \& D. Milne eds. 
        (Dordrecht: Kluwer), 45

Israel, F. P. et al. 2002, A\&A, submitted

Kennicut, R. C. Jr. \& Hodge, P. W. 1986, ApJ, 306, 130

Lucke, P. B. \& Hodge, P. W. 1970, AJ, 75, 171

Naz\'e, Y., Chu, Y.-H., Points, S. D., Danforth, C. W., Rosado, M., \&
        Chen, C.-H. R. 2001, AJ, 122, 921

Oey, M. S. 1999, in ``New Views of the Magellanic Clouds'', IAU Symposium 
        Vol. 190, Y-H. Chu, N.R. Suntzeff, J.E. Hesser, \& D.A. Bohlender, 
        eds. (Provo: ASP), 78

Parker, J. Wm., Garmany, C. D., Massey, P., \& Walborn, N. R. 1992, 
        AJ, 103, 1205 (PGMW)

Persson, S. E., West, S. C., Carr, D. M., Sivaramakrishnan, A., \& 
        Murphy, D. C. 1992, PASP, 104, 204

Persson, S. E., Murphy, D. C., Krzeminski, W., Roth, M., \& Rieke, M. J.
        1998, AJ, 116, 2475

Rosado, M., Laval, A., Le Coarer, E., Georgelin, Y. P., Amram, P., Marcelin, 
        M., Goldes, G., \& Gach, J. L. 1996, A\&A, 308, 588
  
Rubio, M., Barb\'a, R. H., Walborn, N. R., Probst, R. G., Garc\'{\i}a, J., \&
        Roth, M. R. 1998, AJ, 116, 1708

Scowen, P. A. et al. 1998, AJ, 116, 163

Sinclar, M. W., Carrad, G. J., Caswell, J. L., Norris, R. P., \& 
        Whiteoak, J. B. 1992, MNRAS, 256, 33

Smith, N., Egan, M. P., Carey, S., Price, S. D., Morse, J. A., \& Price, P. A.
        2000, ApJ, 532, L145

Tenoglio-Tagle, G. 1979, A\&A, 71, 79

Vacca, W.D., Garmany, C.D., \& Shull, J.M. 1996, ApJ, 460, 914

Walborn, N.R. \& Parker, J.Wm. 1992, ApJ, 399, L87

Walborn, N. R., Barb\'a, R. H., Brandner, W., Rubio, M., Grebel, E. K., \& 
        Probst, R. G. 1999a, AJ, 117, 225

Walborn, N. R., Drissen, L., Parker, J. Wm., Saha, A., MacKenty, J. W., \&
        White, R. L. 1999b, AJ, 118, 1684

Walborn, N. R., Ma\'{\i}z-Apell\'aniz, J., \& Barb\'a, R. H. 2002, AJ, in press

Walsh, A. J., Burton, M. G., Hyland, A. R., \& Robinson, G. 1999, MNRAS, 905, 
        922

Wang, Q. \& Helfand, D. J. 1991, ApJ, 373, 497

\clearpage

%Table 1

\begin{deluxetable}{rrrrrrrrrrrl}
%\rotate
\tabletypesize{\scriptsize}
\tablecolumns{12}
\tablewidth{0pt}
\tablecaption{IR sources in a field of N11B\label{table1-ir-catalog}}
\tablehead{
\colhead{BRRG \#} & 
\colhead{$\alpha(\rm{J2000})$} &
\colhead{$\delta(\rm{J2000})$} &
\colhead{$J$} &
\colhead{$\sigma(J)$} &
\colhead{$H$} &
\colhead{$\sigma(H)$} &
\colhead{$Ks$} &
\colhead{$\sigma(Ks)$} &
\colhead{$J-H$} &
\colhead{$H-Ks$} &
\colhead{Comment\tablenotemark{1}}
}
\startdata
   1 &  4:56:34.87 & 	$-66:25:04.7$ &  17.58 &	0.11 &	16.90 &  0.08 &	16.86 &  0.07 &	$  0.68 $  &$ 0.04 $  &  4010 \\	
   2 &  4:56:34.98 & 	$-66:25:20.2$ &  15.22 &	0.03 &	15.24 &  0.02 &	15.21 &  0.03 &	$ -0.02 $  &$ 0.03 $  &  3016 \\	
   3 &  4:56:34.98 & 	$-66:25:32.3$ &  16.06 &	0.04 &	15.46 &  0.03 &	15.25 &  0.02 &	$  0.60 $  &$ 0.21 $  &       \\	
   4 &  4:56:35.26 & 	$-66:25:27.8$ &  17.06 &	0.05 &	16.35 &  0.04 &	16.27 &  0.04 &	$  0.71 $  &$ 0.08 $  &       \\	
   5 &  4:56:35.41 & 	$-66:24:15.3$ &  16.69 &	0.04 &	16.42 &  0.06 &	16.37 &  0.07 &	$  0.27 $  &$ 0.05 $  &  3019 \\	
   6 &  4:56:35.51 & 	$-66:25:19.3$ &  17.60 &	0.06 &	17.03 &  0.07 &	16.84 &  0.07 &	$  0.57 $  &$ 0.19 $  &       \\	
   7 &  4:56:35.94 & 	$-66:25:05.1$ &  18.54 &	0.10 &	17.55 &  0.09 &	16.61 &  0.09 &	$  0.99 $  &$ 0.94 $  &       \\	
   8 &  4:56:35.98 & 	$-66:24:20.2$ &  16.33 &	0.04 &	15.61 &  0.02 &	15.40 &  0.04 &	$  0.72 $  &$ 0.21 $  &  4011 \\	
   9 &  4:56:36.14 & 	$-66:24:21.9$ &  17.00 &	0.05 &	15.95 &  0.05 &	15.30 &  0.05 &	$  1.05 $  &$ 0.65 $  &       \\	
  10 &  4:56:36.67 & 	$-66:24:13.2$ &  15.86 &	0.07 &	15.63 &  0.06 &	15.57 &  0.08 &	$  0.23 $  &$ 0.06 $  &  3025, 3027 \\	
  11 &  4:56:36.74 & 	$-66:25:22.7$ &  18.53 &	0.06 &\nodata &\nodata&\nodata&\nodata&\nodata     &\nodata   &  3026 \\	
  12 &  4:56:37.09 & 	$-66:25:10.4$ &  17.52 &	0.04 &	16.98 &  0.04 &	16.90 &  0.05 &	$  0.54 $  &$ 0.08 $  &       \\	
  13 &  4:56:37.17 & 	$-66:24:43.8$ &  16.42 &	0.07 &	15.56 &  0.04 &	14.81 &  0.05 &	$  0.86 $  &$ 0.75 $  &  3029 \\	
  14 &  4:56:37.25 & 	$-66:24:37.0$ &  17.79 &	0.05 &	17.75 &  0.08 &	17.67 &  0.10 &	$  0.04 $  &$ 0.08 $  &  3030 \\	
  15 &  4:56:37.28 & 	$-66:24:21.1$ &\nodata &      \nodata&	17.88 &  0.10 &	17.72 &  0.15 &\nodata     &$ 0.16 $  &  3031 \\	
  16 &  4:56:37.28 & 	$-66:25:15.7$ &  18.56 &	0.07 &	17.89 &  0.06 &	17.70 &  0.11 &	$  0.67 $  &$ 0.19 $  &       \\	
  17 &  4:56:37.58 & 	$-66:24:42.8$ &  18.65 &	0.10 &	18.25 &  0.11 &\nodata&\nodata&	$  0.40 $  &\nodata   &       \\	
  18 &  4:56:37.76 & 	$-66:24:29.2$ &  18.43 &	0.10 &\nodata &\nodata&\nodata&\nodata&\nodata     &\nodata   &  3032 \\
  19 &  4:56:38.29 & 	$-66:25:45.5$ &  15.17 &	0.01 &	14.60 &  0.02 &	14.44 &  0.02 &	$  0.57 $  &$ 0.16 $  &  4013 \\	
  20 &  4:56:38.38 & 	$-66:25:40.7$ &  17.51 &	0.05 &	17.09 &  0.05 &	16.92 &  0.08 &	$  0.42 $  &$ 0.17 $  &       \\	
  21 &  4:56:38.40 & 	$-66:24:33.7$ &\nodata &      \nodata&	18.24 &  0.08 &\nodata&\nodata&\nodata	   &\nodata   &       \\	
  22 &  4:56:38.41 & 	$-66:24:43.5$ &  18.58 &	0.07 &	17.56 &  0.06 &	17.13 &  0.07 &	$  1.02 $  &$ 0.43 $  &       \\	
  23 &  4:56:38.48 & 	$-66:24:17.4$ &  17.49 &	0.06 &	16.98 &  0.04 &	16.79 &  0.08 &	$  0.51 $  &$ 0.19 $  &       \\	
  24 &  4:56:38.82 & 	$-66:24:51.1$ &  18.33 &	0.08 &	18.37 &  0.12 &\nodata&\nodata&	$ -0.04 $  &\nodata   &  3037 \\	
  25 &  4:56:38.98 & 	$-66:25:51.3$ &  17.21 &	0.04 &	17.14 &  0.06 &	17.05 &  0.08 &	$  0.07 $  &$ 0.09 $  &  3039 \\	
  26 &  4:56:39.02 & 	$-66:24:45.2$ &  16.53 &	0.08 &	16.07 &  0.10 &	15.88 &  0.10 &	$  0.46 $  &$ 0.19 $  &  3040 \\	
  27 &  4:56:39.24 & 	$-66:24:49.7$ &  14.99 &	0.02 &	15.05 &  0.02 &	15.07 &  0.03 &	$ -0.06 $  &$-0.02 $  &  3042 \\	
  28 &  4:56:39.32 & 	$-66:24:52.1$ &  17.65 &	0.05 &	17.60 &  0.07 &	17.73 &  0.09 &	$  0.05 $  &$-0.13 $  &  3043 \\	
  29 &  4:56:39.53 & 	$-66:25:30.0$ &  17.66 &	0.06 &	17.57 &  0.06 &	17.63 &  0.10 &	$  0.09 $  &$-0.06 $  &  3044 \\	
  30 &  4:56:39.55 & 	$-66:24:49.2$ &  18.28 &	0.08 &	17.84 &  0.07 &	17.62 &  0.07 &	$  0.44 $  &$ 0.22 $  &       \\	
  31 &  4:56:39.84 & 	$-66:24:54.1$ &  15.35 &	0.01 &	15.43 &  0.01 &	15.42 &  0.03 &	$ -0.08 $  &$ 0.01 $  &  3045 \\	
  32 &  4:56:39.88 & 	$-66:25:51.4$ &  17.43 &	0.07 &	17.59 &  0.11 &	17.61 &  0.13 &	$ -0.16 $  &$-0.02 $  &  3046 \\	
  33 &  4:56:39.93 & 	$-66:25:07.6$ &  17.75 &	0.05 &	17.85 &  0.06 &	17.74 &  0.09 &	$ -0.10 $  &$ 0.11 $  &  3047 \\	
  34 &  4:56:39.98 & 	$-66:25:48.9$ &  17.92 &	0.06 &	17.83 &  0.08 &	17.88 &  0.13 &	$  0.09 $  &$-0.05 $  &  3048 \\	
  35 &  4:56:40.05 & 	$-66:24:29.8$ &  17.31 &	0.04 &	16.76 &  0.04 &	16.71 &  0.06 &	$  0.55 $  &$ 0.05 $  &  4015 \\	
  36 &  4:56:40.68 & 	$-66:24:50.5$ &  16.46 &	0.03 &	16.54 &  0.04 &	16.53 &  0.05 &	$ -0.08 $  &$ 0.01 $  &  3051 \\	
  37 &  4:56:40.84 & 	$-66:25:29.1$ &  17.81 &	0.05 &	17.33 &  0.04 &	17.20 &  0.07 &	$  0.48 $  &$ 0.13 $  &       \\	
  38 &  4:56:41.06 & 	$-66:24:47.4$ &  17.36 &	0.04 &	16.73 &  0.04 &	16.47 &  0.04 &	$  0.63 $  &$ 0.26 $  &       \\	
  39 &  4:56:41.10 & 	$-66:24:40.2$ &  13.32 &	0.01 &	13.33 &  0.01 &	13.34 &  0.02 &	$ -0.01 $  &$-0.01 $  &  3053 \\	
  40 &  4:56:41.39 & 	$-66:25:10.9$ &  18.34 &	0.10 &	18.01 &  0.09 &	17.77 &  0.12 &	$  0.33 $  &$ 0.24 $  &       \\	
  41 &  4:56:41.45 & 	$-66:24:27.1$ &  18.37 &	0.07 &	18.22 &  0.10 &\nodata&\nodata&	$  0.15 $  &\nodata   &  3055 \\	
  42 &  4:56:41.77 & 	$-66:24:57.4$ &  17.59 &	0.07 &	17.11 &  0.05 &	16.90 &  0.06 &	$  0.48 $  &$ 0.21 $  &       \\	
  43 &  4:56:41.84 & 	$-66:24:32.5$ &  18.46 &	0.06 &	18.29 &  0.09 &\nodata&\nodata&	$  0.17 $  &\nodata   &  4018 \\	
  44 &  4:56:42.02 & 	$-66:25:05.0$ &  17.96 &	0.07 &	17.71 &  0.06 &	17.55 &  0.10 &	$  0.25 $  &$ 0.16 $  &  3056 \\	
  45 &  4:56:42.08 & 	$-66:24:58.2$ &  17.90 &	0.06 &	17.52 &  0.07 &	17.26 &  0.07 &	$  0.38 $  &$ 0.26 $  &       \\	
  46 &  4:56:42.08 & 	$-66:25:51.7$ &\nodata &      \nodata&	17.64 &  0.11 &	17.55 &  0.17 &	\nodata    &$ 0.09 $  &       \\	
  47 &  4:56:42.09 & 	$-66:25:36.7$ &  17.39 &	0.04 &	17.41 &  0.06 &	17.49 &  0.08 &	$ -0.02 $  &$-0.08 $  &  3057 \\	
  48 &  4:56:42.16 & 	$-66:24:54.2$ &  14.27 &	0.01 &	14.27 &  0.02 &	14.23 &  0.02 &	$  0.00 $  &$ 0.04 $  &  3058 \\	
  49 &  4:56:42.30 & 	$-66:25:17.3$ &  15.10 &	0.02 &	14.71 &  0.02 &	14.59 &  0.02 &	$  0.39 $  &$ 0.12 $  &  4020 \\	
  50 &  4:56:42.40 & 	$-66:24:57.2$ &  17.80 &	0.06 &	17.62 &  0.06 &	17.09 &  0.08 &	$  0.18 $  &$ 0.53 $  &       \\	
  51 &  4:56:42.42 & 	$-66:24:34.4$ &\nodata &      \nodata&	18.13 &  0.08 &\nodata&\nodata&	\nodata    &\nodata   &  4023 \\	
  52 &  4:56:42.42 & 	$-66:24:14.9$ &  18.12 &	0.08 &	17.47 &  0.06 &	17.30 &  0.10 &	$  0.65 $  &$ 0.17 $  &       \\	
  53 &  4:56:42.43 & 	$-66:24:53.5$ &  16.48 &	0.03 &	16.55 &  0.03 &	16.44 &  0.05 &	$ -0.07 $  &$ 0.11 $  &  3059 \\	
  54 &  4:56:42.46 & 	$-66:25:34.8$ &  17.85 &	0.05 &	17.33 &  0.04 &	17.16 &  0.07 &	$  0.52 $  &$ 0.17 $  &       \\	
  55 &  4:56:42.51 & 	$-66:24:44.3$ &  17.97 &	0.07 &	17.41 &  0.06 &	17.34 &  0.10 &	$  0.56 $  &$ 0.07 $  &  4022 at $1''$\\	
  56 &  4:56:42.52 & 	$-66:25:17.6$ &  13.69 &	0.01 &	13.72 &  0.01 &	13.71 &  0.02 &	$ -0.03 $  &$ 0.01 $  &  3061 \\	
  57 &  4:56:42.62 & 	$-66:25:05.7$ &  18.42 &	0.08 &	17.76 &  0.07 &	17.72 &  0.10 &	$  0.66 $  &$ 0.04 $  &       \\	
  58 &  4:56:42.65 & 	$-66:24:39.1$ &  17.93 &	0.05 &	17.79 &  0.07 &	17.84 &  0.09 &	$  0.14 $  &$-0.05 $  &  3062 \\	
  59 &  4:56:42.79 & 	$-66:24:25.9$ &  17.75 &	0.10 &	17.22 &  0.06 &	16.99 &  0.07 &	$  0.53 $  &$ 0.23 $  &  4025 at $1\farcs3$\\	
  60 &  4:56:42.79 & 	$-66:24:35.8$ &  18.37 &	0.08 &\nodata &\nodata&\nodata&\nodata&	\nodata    &\nodata   &  4024 \\	
  61 &  4:56:42.79 & 	$-66:25:02.7$ &  15.42 &	0.04 &	15.38 &  0.04 &	15.36 &  0.03 &	$  0.04 $  &$ 0.02 $  &  3063 \\	
  62 &  4:56:42.85 & 	$-66:24:58.8$ &  18.04 &	0.08 &	17.79 &  0.09 &	17.54 &  0.12 &	$  0.25 $  &$ 0.25 $  &       \\	
  63 &  4:56:42.86 & 	$-66:24:40.9$ &  18.49 &	0.08 &	18.16 &  0.06 &\nodata&\nodata&	$  0.33 $  &\nodata   &       \\	
  64 &  4:56:42.94 & 	$-66:25:01.0$ &  16.55 &	0.13 &	16.48 &  0.12 &	16.37 &  0.14 &	$  0.07 $  &$ 0.11 $  &  3064 \\	
  65 &  4:56:42.96 & 	$-66:25:04.3$ &  17.71 &	0.06 &	17.50 &  0.09 &	17.06 &  0.07 &	$  0.21 $  &$ 0.44 $  &       \\	
  66 &  4:56:42.98 & 	$-66:25:10.1$ &  18.34 &	0.05 &\nodata &\nodata&\nodata&\nodata&	\nodata    & \nodata  &  4026 \\	
  67 &  4:56:43.04 & 	$-66:25:02.6$ &  13.86 &	0.03 &	13.80 &  0.03 &	13.78 &  0.03 &	$  0.06 $  &$ 0.02 $  &  3065 \\	
  68 &  4:56:43.14 & 	$-66:25:08.5$ &  17.86 &	0.06 &	17.82 &  0.06 &	17.71 &  0.10 &	$  0.04 $  &$ 0.11 $  &  3067 \\	
  69 &  4:56:43.15 & 	$-66:24:14.9$ &  17.93 &	0.07 &	17.13 &  0.05 &	16.77 &  0.07 &	$  0.80 $  &$ 0.36 $  &       \\	
  70 &  4:56:43.19 & 	$-66:24:37.8$ &  15.01 &	0.02 &	14.24 &  0.01 &	14.07 &  0.02 &	$  0.77 $  &$ 0.17 $  &  4027 \\	
  71 &  4:56:43.21 & 	$-66:24:34.4$ &  18.29 &	0.07 &	18.23 &  0.09 &\nodata&\nodata&	$  0.06 $  & \nodata  &  3068 \\	
  72 &  4:56:43.23 & 	$-66:24:54.2$ &  16.28 &	0.09 &	16.17 &  0.09 &	16.35 &  0.09 &	$  0.11 $  &$-0.18 $  &  3069 \\	
  73 &  4:56:43.28 & 	$-66:25:05.5$ &  17.55 &	0.07 &	17.25 &  0.08 &	17.07 &  0.09 &	$  0.30 $  &$ 0.18 $  &       \\	
  74 &  4:56:43.29 & 	$-66:24:57.7$ &  18.09 &	0.06 &	17.60 &  0.07 &	17.24 &  0.08 &	$  0.49 $  &$ 0.36 $  &       \\	
  75 &  4:56:43.29 & 	$-66:25:02.0$ &  12.65 &	0.04 &	12.63 &  0.04 &	12.65 &  0.04 &	$  0.02 $  &$-0.02 $  &  3070, Multiple star\\	
  76 &  4:56:43.30 & 	$-66:25:21.0$ &  16.54 &	0.02 &	16.03 &  0.03 &	15.84 &  0.03 &	$  0.51 $  &$ 0.19 $  &  4028 \\	
  77 &  4:56:43.31 & 	$-66:25:33.9$ &  16.88 &	0.03 &	16.84 &  0.03 &	16.79 &  0.04 &	$  0.04 $  &$ 0.05 $  &  3071 \\	
  78 &  4:56:43.34 & 	$-66:24:59.6$ &  17.06 &	0.07 &	16.89 &  0.07 &	16.87 &  0.12 &	$  0.17 $  &$ 0.02 $  &  4029 \\	
  79 &  4:56:43.37 & 	$-66:24:43.7$ &  18.53 &	0.09 &	18.32 &  0.10 &\nodata&\nodata&	$  0.21 $  & \nodata  &  3072 \\	
  80 &  4:56:43.38 & 	$-66:24:54.1$ &  14.85 &	0.05 &	14.88 &  0.05 &	14.82 &  0.05 &	$ -0.03 $  &$ 0.06 $  &  3073, 3075 at $1''$\\	
  81 &  4:56:43.46 & 	$-66:25:04.8$ &  18.20 &	0.10 &	18.02 &  0.11 &\nodata&\nodata&	$  0.18 $  & \nodata  &       \\	
  82 &  4:56:43.74 & 	$-66:25:03.6$ &  17.64 &	0.06 &	17.40 &  0.08 &	17.43 &  0.10 &	$  0.24 $  &$-0.03 $  &  4031 \\	
  83 &  4:56:43.75 & 	$-66:25:26.7$ &  17.44 &	0.04 &	16.91 &  0.04 &	16.60 &  0.05 &	$  0.53 $  &$ 0.31 $  &       \\	
  84 &  4:56:43.77 & 	$-66:24:35.6$ &  16.86 &	0.03 &	16.79 &  0.04 &	16.71 &  0.05 &	$  0.07 $  &$ 0.08 $  &  3078 \\	
  85 &  4:56:43.77 & 	$-66:25:21.3$ &  16.51 &	0.03 &	16.52 &  0.03 &	16.50 &  0.05 &	$ -0.01 $  &$ 0.02 $  &  3077 \\	
  86 &  4:56:43.79 & 	$-66:25:02.2$ &  16.68 &	0.09 &	16.83 &  0.09 &	16.60 &  0.11 &	$ -0.15 $  &$ 0.23 $  &  3076 \\	
  87 &  4:56:43.81 & 	$-66:24:59.2$ &  18.19 &	0.10 &	17.85 &  0.09 &	17.60 &  0.11 &	$  0.34 $  &$ 0.25 $  &       \\	
  88 &  4:56:43.89 & 	$-66:24:14.6$ &  16.58 &	0.04 &	16.39 &  0.04 &	16.15 &  0.04 &	$  0.19 $  &$ 0.24 $  &  3079 \\	
  89 &  4:56:43.91 & 	$-66:25:13.7$ &  18.50 &	0.06 &	17.78 &  0.06 &	17.01 &  0.06 &	$  0.72 $  &$ 0.77 $  &       \\	
  90 &  4:56:43.93 & 	$-66:24:30.4$ &  18.33 &	0.07 &	17.72 &  0.05 &	17.40 &  0.08 &	$  0.61 $  &$ 0.32 $  &       \\	
  91 &  4:56:43.95 & 	$-66:24:53.6$ &  17.22 &	0.04 &	17.37 &  0.07 &	17.19 &  0.05 &	$ -0.15 $  &$ 0.18 $  &  3080 \\	
  92 &  4:56:44.06 & 	$-66:25:06.9$ &  15.64 &	0.05 &	15.66 &  0.03 &	15.60 &  0.05 &	$ -0.02 $  &$ 0.06 $  &  3081, 3083 at $1''$\\	
  93 &  4:56:44.11 & 	$-66:24:46.3$ &  16.90 &	0.05 &	16.92 &  0.04 &	16.95 &  0.06 &	$ -0.02 $  &$-0.03 $  &  3082 \\	
  94 &  4:56:44.15 & 	$-66:24:53.0$ &  17.31 &	0.06 &	17.12 &  0.07 &	16.97 &  0.08 &	$  0.19 $  &$ 0.15 $  &  3084 \\	
  95 &  4:56:44.19 & 	$-66:25:32.8$ &  17.69 &	0.04 &	17.26 &  0.03 &	17.03 &  0.05 &	$  0.43 $  &$ 0.23 $  &       \\	
  96 &  4:56:44.20 & 	$-66:24:32.4$ &  17.63 &	0.06 &	17.03 &  0.05 &	16.95 &  0.07 &	$  0.60 $  &$ 0.08 $  &       \\	
  97 &  4:56:44.27 & 	$-66:24:56.9$ &  18.58 &	0.12 &\nodata &\nodata&\nodata&\nodata&	 \nodata   & \nodata  &  3085 at $1\farcs1$\\	
  98 &  4:56:44.41 & 	$-66:25:00.3$ &  16.38 &	0.05 &	16.35 &  0.06 &	16.38 &  0.07 &	$  0.03 $  &$-0.03 $  &  3087 \\	
  99 &  4:56:44.45 & 	$-66:24:52.8$ &  18.47 &	0.13 &\nodata &\nodata&\nodata&\nodata&	 \nodata   & \nodata  &       \\
 100 &  4:56:44.47 & 	$-66:24:15.6$ &  17.68 &	0.06 &	17.53 &  0.09 &	17.52 &  0.09 &	$  0.15 $  &$ 0.01 $  &  3091 \\	
 101 &  4:56:44.48 & 	$-66:24:55.6$ &  15.93 &	0.05 &	15.91 &  0.06 &	15.93 &  0.04 &	$  0.02 $  &$-0.02 $  &  3088 \\	
 102 &  4:56:44.58 & 	$-66:24:33.4$ &  14.85 &	0.01 &	14.87 &  0.02 &	14.84 &  0.02 &	$ -0.02 $  &$ 0.03 $  &  3089 \\	
 103 &  4:56:44.61 & 	$-66:24:56.6$ &  15.24 &	0.01 &	15.30 &  0.01 &	15.24 &  0.02 &	$ -0.06 $  &$ 0.06 $  &  3090 \\	
 104 &  4:56:44.63 & 	$-66:24:59.5$ &  15.94 &	0.04 &	15.97 &  0.05 &	15.90 &  0.04 &	$ -0.03 $  &$ 0.07 $  &  3092 \\	
 105 &  4:56:44.68 & 	$-66:24:54.7$ &  15.68 &	0.02 &	15.58 &  0.01 &	15.43 &  0.03 &	$  0.10 $  &$ 0.15 $  &  3093 \\	
 106 &  4:56:44.72 & 	$-66:24:50.7$ &  16.17 &	0.03 &	16.14 &  0.03 &	16.19 &  0.04 &	$  0.03 $  &$-0.05 $  &  3095 \\	
 107 &  4:56:44.74 & 	$-66:24:35.5$ &  18.45 &	0.11 &	18.18 &  0.07 &\nodata&\nodata&	$  0.27 $  & \nodata  &       \\	
 108 &  4:56:44.76 & 	$-66:24:29.2$ &  18.48 &	0.10 &	17.22 &  0.06 &	16.02 &  0.06 &	$  1.26 $  &$ 1.20 $  &       \\	
 109 &  4:56:44.79 & 	$-66:25:02.4$ &  17.22 &	0.04 &	16.72 &  0.05 &	16.63 &  0.05 &	$  0.50 $  &$ 0.09 $  &  4034 \\	
 110 &  4:56:44.80 & 	$-66:24:43.3$ &\nodata &      \nodata&	18.24 &  0.07 &\nodata&\nodata&	 \nodata   & \nodata  &       \\	
 111 &  4:56:44.81 & 	$-66:25:07.1$ &  15.32 &	0.05 &	15.33 &  0.04 &	15.37 &  0.04 &	$ -0.01 $  &$-0.04 $  &  3097, 3096 at $1''$\\	
 112 &  4:56:44.86 & 	$-66:25:44.0$ &  17.43 &	0.06 &	17.51 &  0.07 &	17.50 &  0.08 &	$ -0.08 $  &$ 0.01 $  &  3099 \\	
 113 &  4:56:44.93 & 	$-66:24:29.4$ &\nodata &      \nodata&	17.52 &  0.08 &	16.75 &  0.09 &	$ 81.38 $  &$ 0.77 $  &       \\	
 114 &  4:56:44.95 & 	$-66:25:04.8$ &  18.26 &	0.06 &	18.06 &  0.08 &\nodata&\nodata&	$  0.20 $  & \nodata  &       \\	
 115 &  4:56:44.96 & 	$-66:25:19.2$ &  17.40 &	0.04 &	16.94 &  0.03 &	16.74 &  0.05 &	$  0.46 $  &$ 0.20 $  &       \\	
 116 &  4:56:45.06 & 	$-66:24:47.3$ &  16.05 &	0.03 &	15.40 &  0.01 &	15.16 &  0.03 &	$  0.65 $  &$ 0.24 $  &       \\	
 117 &  4:56:45.09 & 	$-66:24:57.9$ &  16.67 &	0.05 &	16.50 &  0.07 &	16.62 &  0.06 &	$  0.17 $  &$-0.12 $  &  3101 \\	
 118 &  4:56:45.11 & 	$-66:24:49.0$ &  18.55 &	0.08 &\nodata &\nodata&\nodata&\nodata&	 \nodata   & \nodata  &       \\
 119 &  4:56:45.22 & 	$-66:25:10.2$ &  13.70 &	0.01 &	13.69 &  0.01 &	13.67 &  0.02 &	$  0.01 $  &$ 0.02 $  &  3100 \\	
 120 &  4:56:45.24 & 	$-66:24:50.9$ &  18.06 &	0.06 &	17.70 &  0.06 &	17.47 &  0.06 &	$  0.36 $  &$ 0.23 $  &       \\	
 121 &  4:56:45.38 & 	$-66:24:37.5$ &  16.03 &	0.04 &	16.05 &  0.02 &	16.06 &  0.04 &	$ -0.02 $  &$-0.01 $  &  3103 \\	
 122 &  4:56:45.39 & 	$-66:24:46.0$ &  13.76 &	0.01 &	13.77 &  0.01 &	13.80 &  0.02 &	$ -0.01 $  &$-0.03 $  &  3102 \\	
 123 &  4:56:45.41 & 	$-66:24:26.7$ &  17.79 &	0.08 &	17.30 &  0.07 &	17.12 &  0.10 &	$  0.49 $  &$ 0.18 $  &  3104 \\	
 124 &  4:56:45.45 & 	$-66:25:15.5$ &  18.37 &	0.10 &\nodata &\nodata&\nodata&\nodata&	 \nodata   & \nodata  &  3105 \\
 125 &  4:56:45.57 & 	$-66:24:39.4$ &  16.93 &	0.03 &	16.92 &  0.05 &	17.04 &  0.07 &	$  0.01 $  &$-0.12 $  &  3106 \\	
 126 &  4:56:45.64 & 	$-66:25:08.5$ &  17.46 &	0.06 &	17.49 &  0.06 &	17.58 &  0.08 &	$ -0.03 $  &$-0.09 $  &  3107 \\	
 127 &  4:56:45.82 & 	$-66:24:29.9$ &  17.17 &	0.12 &	16.64 &  0.10 &	16.24 &  0.08 &	$  0.53 $  &$ 0.40 $  &  3108 at $0\farcs8$\\	
 128 &  4:56:45.87 & 	$-66:25:36.9$ &  17.10 &	0.03 &	16.59 &  0.04 &	16.36 &  0.05 &	$  0.51 $  &$ 0.23 $  &       \\	
 129 &  4:56:45.96 & 	$-66:24:21.8$ &\nodata &      \nodata&	17.85 &  0.10 &	17.07 &  0.09 &	 \nodata   &$ 0.78 $  &       \\	
 130 &  4:56:46.06 & 	$-66:24:49.3$ &  16.44 &	0.03 &	16.41 &  0.03 &	16.38 &  0.05 &	$  0.03 $  &$ 0.03 $  &  3109 \\	
 131 &  4:56:46.11 & 	$-66:24:52.7$ &  17.21 &	0.05 &	17.19 &  0.06 &	17.00 &  0.08 &	$  0.02 $  &$ 0.19 $  &  3110 \\	
 132 &  4:56:46.33 & 	$-66:24:46.7$ &  17.19 &	0.09 &	17.12 &  0.09 &	17.03 &  0.06 &	$  0.07 $  &$ 0.09 $  &  3113 \\	
 133 &  4:56:46.36 & 	$-66:25:08.3$ &  17.72 &	0.05 &	17.68 &  0.05 &	17.50 &  0.10 &	$  0.04 $  &$ 0.18 $  &  3116 \\	
 134 &  4:56:46.49 & 	$-66:24:45.8$ &  17.86 &	0.15 &	17.39 &  0.12 &	16.99 &  0.10 &	$  0.47 $  &$ 0.40 $  &       \\	
 135 &  4:56:46.55 & 	$-66:24:51.3$ &  15.18 &	0.03 &	15.12 &  0.03 &	15.15 &  0.04 &	$  0.06 $  &$-0.03 $  &  3115 \\	
 136 &  4:56:46.58 & 	$-66:24:14.7$ &  17.09 &	0.04 &	16.38 &  0.04 &	16.25 &  0.05 &	$  0.71 $  &$ 0.13 $  &       \\	
 137 &  4:56:46.72 & 	$-66:25:09.2$ &  17.85 &	0.06 &	17.83 &  0.06 &	17.61 &  0.09 &	$  0.02 $  &$ 0.22 $  &  3118 \\	
 138 &  4:56:46.75 & 	$-66:24:53.0$ &  18.07 &	0.06 &	17.59 &  0.05 &	17.34 &  0.09 &	$  0.48 $  &$ 0.25 $  &       \\	
 139 &  4:56:46.78 & 	$-66:24:43.2$ &  17.10 &	0.07 &	16.90 &  0.05 &	16.95 &  0.09 &	$  0.20 $  &$-0.05 $  &  3121 \\	
 140 &  4:56:46.83 & 	$-66:24:46.4$ &  12.92 &	0.03 &	12.96 &  0.03 &	12.97 &  0.04 &	$ -0.04 $  &$-0.01 $  &  3120 \\	
 141 &  4:56:46.89 & 	$-66:25:44.1$ &  17.92 &	0.05 &	18.07 &  0.07 &	17.70 &  0.11 &	$ -0.15 $  &$ 0.37 $  &  3122 \\	
 142 &  4:56:47.01 & 	$-66:24:28.3$ &  15.82 &	0.03 &	15.25 &  0.01 &	15.06 &  0.05 &	$  0.57 $  &$ 0.19 $  &  4038 \\	
 143 &  4:56:47.02 & 	$-66:24:55.8$ &  18.41 &	0.08 &	17.97 &  0.06 &	17.63 &  0.09 &	$  0.44 $  &$ 0.34 $  &       \\	
 144 &  4:56:47.03 & 	$-66:24:31.3$ &  14.78 &	0.02 &	14.72 &  0.02 &	14.64 &  0.05 &	$  0.06 $  &$ 0.08 $  &  3123 \\	
 145 &  4:56:47.07 & 	$-66:25:02.6$ &  16.27 &	0.03 &	16.20 &  0.03 &	16.11 &  0.03 &	$  0.07 $  &$ 0.09 $  &  3127 \\	
 146 &  4:56:47.11 & 	$-66:24:58.9$ &  14.10 &	0.02 &	14.08 &  0.02 &	14.05 &  0.02 &	$  0.02 $  &$ 0.03 $  &  3126 \\	
 147 &  4:56:47.13 & 	$-66:24:30.5$ &  17.56 &	0.11 &	16.70 &  0.08 &	15.48 &  0.09 &	$  0.86 $  &$ 1.22 $  &       \\	
 148 &  4:56:47.17 & 	$-66:24:33.1$ &\nodata &      \nodata&	17.92 &  0.11 &	16.72 &  0.12 &	 \nodata   &$ 1.20 $  &       \\	
 149 &  4:56:47.23 & 	$-66:24:52.1$ &  16.48 &	0.07 &	16.41 &  0.07 &	16.43 &  0.09 &	$  0.07 $  &$-0.02 $  &  3129 \\	
 150 &  4:56:47.25 & 	$-66:24:41.9$ &  14.87 &	0.02 &	14.90 &  0.01 &	14.95 &  0.02 &	$ -0.03 $  &$-0.05 $  &  3128 \\	
 151 &  4:56:47.29 & 	$-66:25:11.3$ &  17.69 &	0.05 &	17.59 &  0.05 &	17.53 &  0.08 &	$  0.10 $  &$ 0.06 $  &  3130 \\	
 152 &  4:56:47.33 & 	$-66:24:37.2$ &  17.33 &	0.15 &	17.16 &  0.12 &	16.42 &  0.16 &	$  0.17 $  &$ 0.74 $  &  3131 at $0\farcs7$, 4039 at $1\farcs3$\\	
 153 &  4:56:47.39 & 	$-66:25:18.1$ &  17.35 &	0.04 &	16.90 &  0.03 &	16.75 &  0.05 &	$  0.45 $  &$ 0.15 $  &       \\	
 154 &  4:56:47.49 & 	$-66:24:54.1$ &  18.46 &	0.07 &\nodata &\nodata&\nodata&\nodata&	 \nodata   & \nodata  &       \\
 155 &  4:56:47.50 & 	$-66:24:57.1$ &  16.23 &	0.03 &	16.25 &  0.03 &	16.22 &  0.04 &	$ -0.02 $  &$ 0.03 $  &  3134 \\	
 156 &  4:56:47.53 & 	$-66:24:56.0$ &  16.82 &	0.03 &	16.83 &  0.04 &	16.76 &  0.05 &	$ -0.01 $  &$ 0.07 $  &  3135 \\	
 157 &  4:56:47.54 & 	$-66:24:33.5$ &  17.05 &	0.05 &	16.25 &  0.05 &	15.15 &  0.05 &	$  0.80 $  &$ 1.10 $  &  3133 \\	
 158 &  4:56:47.55 & 	$-66:24:50.1$ &  18.07 &	0.11 &	17.91 &  0.11 &	17.71 &  0.11 &	$  0.16 $  &$ 0.20 $  &  3137 \\	
 159 &  4:56:47.56 & 	$-66:24:44.3$ &  18.34 &	0.07 &\nodata &\nodata&\nodata&\nodata&	 \nodata   & \nodata  &  3136 \\
 160 &  4:56:47.61 & 	$-66:25:14.1$ &  17.33 &	0.04 &	16.87 &  0.03 &	16.74 &  0.07 &	$  0.46 $  &$ 0.13 $  &       \\	
 161 &  4:56:47.64 & 	$-66:25:08.0$ &  17.88 &	0.05 &	17.19 &  0.04 &	16.84 &  0.06 &	$  0.69 $  &$ 0.35 $  &       \\	
 162 &  4:56:47.74 & 	$-66:24:33.8$ &  16.83 &	0.06 &	16.18 &  0.06 &	15.60 &  0.09 &	$  0.65 $  &$ 0.58 $  &       \\	
 163 &  4:56:47.77 & 	$-66:24:17.4$ &  17.20 &	0.05 &	16.55 &  0.14 &	16.33 &  0.07 &	$  0.65 $  &$ 0.22 $  &       \\	
 164 &  4:56:47.78 & 	$-66:24:52.1$ &  18.58 &	0.06 &\nodata &\nodata&\nodata&\nodata&	 \nodata   & \nodata  &       \\
 165 &  4:56:47.88 & 	$-66:24:35.6$ &  16.84 &	0.12 &	16.72 &  0.12 &	16.35 &  0.14 &	$  0.12 $  &$ 0.37 $  &  3138 \\	
 166 &  4:56:47.89 & 	$-66:24:46.1$ &  18.14 &	0.07 &\nodata &\nodata&\nodata&\nodata&	 \nodata   & \nodata  &  3139 \\
 167 &  4:56:48.05 & 	$-66:24:54.7$ &  17.47 &	0.07 &	17.33 &  0.04 &	17.28 &  0.08 &	$  0.14 $  &$ 0.05 $  &  3142 \\	
 168 &  4:56:48.10 & 	$-66:25:49.0$ &  14.57 &	0.01 &	13.88 &  0.02 &	13.64 &  0.02 &	$  0.69 $  &$ 0.24 $  &  3143 \\	
 169 &  4:56:48.39 & 	$-66:25:21.3$ &  18.55 &	0.09 &	18.15 &  0.08 &	17.50 &  0.07 &	$  0.40 $  &$ 0.65 $  &       \\	
 170 &  4:56:48.47 & 	$-66:24:48.9$ &  17.80 &	0.06 &	17.75 &  0.09 &	17.65 &  0.11 &	$  0.05 $  &$ 0.10 $  &  3145 \\	
 171 &  4:56:48.47 & 	$-66:24:42.8$ &  16.52 &	0.03 &	16.62 &  0.03 &	16.72 &  0.05 &	$ -0.10 $  &$-0.10 $  &  3146 \\	
 172 &  4:56:48.63 & 	$-66:25:11.5$ &  16.97 &	0.04 &	16.96 &  0.04 &	16.99 &  0.07 &	$  0.01 $  &$-0.03 $  &  3147 \\	
 173 &  4:56:48.76 & 	$-66:24:15.5$ &  15.37 &	0.03 &	15.39 &  0.02 &	15.39 &  0.03 &	$ -0.02 $  &$ 0.00 $  &  3148 \\	
 174 &  4:56:48.78 & 	$-66:24:37.5$ &  18.36 &	0.09 &	17.71 &  0.09 &	17.36 &  0.12 &	$  0.65 $  &$ 0.35 $  &       \\	
 175 &  4:56:48.78 & 	$-66:24:34.1$ &  18.62 &	0.12 &	17.91 &  0.08 &\nodata&\nodata&	$  0.71 $  & \nodata  &       \\	
 176 &  4:56:48.87 & 	$-66:25:22.2$ &  18.80 &	0.11 &	18.29 &  0.09 &	17.38 &  0.07 &	$  0.51 $  &$ 0.91 $  &       \\	
 177 &  4:56:48.97 & 	$-66:24:46.8$ &  17.08 &	0.04 &	17.12 &  0.04 &	17.02 &  0.07 &	$ -0.04 $  &$ 0.10 $  &  3150 \\	
 178 &  4:56:49.01 & 	$-66:24:18.0$ &  17.92 &	0.09 &	17.79 &  0.09 &	17.80 &  0.15 &	$  0.13 $  &$-0.01 $  &  3151 \\	
 179 &  4:56:49.07 & 	$-66:24:44.8$ &  17.61 &	0.05 &	17.74 &  0.06 &	17.61 &  0.10 &	$ -0.13 $  &$ 0.13 $  &  3152 \\	
 180 &  4:56:49.93 & 	$-66:24:37.1$ &  18.24 &	0.11 &\nodata &\nodata&\nodata&\nodata&  \nodata   & \nodata  &       \\
 181 &  4:56:50.05 & 	$-66:24:44.2$ &  17.75 &	0.05 &	17.17 &  0.04 &	17.19 &  0.09 &	$  0.58 $  &$-0.02 $  &       \\	
 182 &  4:56:50.59 & 	$-66:24:34.8$ &  12.24 &	0.02 &	12.20 &  0.02 &	12.17 &  0.04 &	$  0.04 $  &$ 0.03 $  &  3157 \\	
 183 &  4:56:50.84 & 	$-66:25:15.3$ &  17.43 &	0.06 &	16.86 &  0.05 &	16.50 &  0.07 &	$  0.57 $  &$ 0.36 $  &       \\	
 184 &  4:56:51.03 & 	$-66:24:51.5$ &  16.29 &	0.04 &	16.28 &  0.04 &	16.32 &  0.05 &	$  0.01 $  &$-0.04 $  &  3159 \\	
\enddata
\tablenotetext{1}{Optical counterpart stars are numbered from Parker et al. 1992}
\end{deluxetable}

\clearpage

%Tabla 2

\begin{deluxetable}{rrrrr}
\tablecolumns{5}
\tablewidth{0pt}
\tablecaption{Comparison of IR photometry of uncrowded stars in LH10
with 2MASS values\label{table2-comp-2MASS}}  
\tablehead{ 
\colhead{2MASS} & 
\colhead{BRRG} &
\colhead{$\Delta(J-J_{\rm 2MASS})$} &
\colhead{$\Delta(H-H_{\rm 2MASS})$} &   
\colhead{$\Delta(Ks-Ks_{\rm 2MASS})$}
}   
\startdata
0456410-662440 &  39 & 0.10 &  0.22 & 0.13\phn \\  
0456452-662510 & 119 & 0.11 &  0.12 & 0.14\phn \\
0456468-662446 & 140 & 0.10 &  0.15 & 0.14\phn \\  
0456506-662435 & 182 & 0.10 &  0.10 & 0.18\phn \\ 
\hline
Mean ($\sigma$)&     & 0.10(0.01) & 0.15 (0.05) & 0.15(0.02) \\ 
\enddata
\end{deluxetable}

\clearpage

%Table 3

\begin{deluxetable}{rrrrr}
\tablecolumns{3}
\tablewidth{0pt}
\tablecaption{Comparison of coordinates of stars in LH10 obtained in
IR images with those values 
published in GSC, Parker et al. 1992, 2MASS, and derived from HST/WFPC2 images.
\label{table3-diff-coord}}  
\tablehead{ 
\colhead{Offset} & 
\colhead{$\Delta\alpha$} &
\colhead{$\Delta\delta$} 
}   
\startdata
GSC--BRRG   & $ 0\fs003\pm0\fs028$ & $ 0\farcs02\pm0\farcs21$\phn \\ 
PGMW--BRRG  & $-0\fs003\pm0\fs010$ & $ 0\farcs22\pm0\farcs05$\phn \\
2MASS--BRRG & $-0\fs013\pm0\fs038$ & $ 0\farcs02\pm0\farcs22$\phn \\
HST--BRRG   & $ 0\fs135\pm0\fs013$ & $-0\farcs70\pm0\farcs14$\phn \\
\enddata
\end{deluxetable}

\clearpage

%Table 4

\begin{deluxetable}{rrrrrrrrrr}
\tabletypesize{\scriptsize}
\tablecolumns{10}
\tablewidth{0pt}
\tablecaption{Candidates Herbig Ae/Be stars in a field of N11B\label{candidates}}
\tablehead{
\colhead{BRRG \#} & 
\colhead{$\alpha(2000)$} &
\colhead{$\delta(2000)$} &
\colhead{$J$} &
\colhead{$\sigma(J)$} &
\colhead{$J-H$} &
\colhead{$\sigma(J-H)$} &
\colhead{$H-Ks$} &
\colhead{$\sigma(H-Ks)$} 
}
\startdata
   7 &  4:56:35.94 & 	$-66:25:05.1$ &  18.54 & 0.10 &	$0.99 $ & 0.14 & $ 0.94$ & 0.13 \\	
   9 &  4:56:36.14 & 	$-66:24:21.9$ &  17.00 & 0.05 &	$1.05 $ & 0.07 & $ 0.65$ & 0.07 \\	
  13 &  4:56:37.17 & 	$-66:24:43.8$ &  16.42 & 0.07 &	$0.86 $ & 0.08 & $ 0.75$ & 0.06 \\	
  26 &  4:56:39.02 &  $-66:24:45.2$ &  16.53 & 0.08 & $0.46 $ & 0.13 & $ 0.19$ & 0.13 \\
  50 &  4:56:42.40 & 	$-66:24:57.2$ &  17.80 & 0.06 &	$0.18 $ & 0.08 & $ 0.53$ & 0.10 \\	
  65 &  4:56:42.96 & 	$-66:25:04.3$ &  17.71 & 0.06 & $0.21 $ & 0.11 & $ 0.44$ & 0.12 \\	
  89 &  4:56:43.91 & 	$-66:25:13.7$ &  18.50 & 0.06 &	$0.72 $ & 0.08 & $ 0.77$ & 0.08 \\	
 108 &  4:56:44.76 & 	$-66:24:29.2$ &  18.48 & 0.10 &	$1.26 $ & 0.12 & $ 1.20$ & 0.12 \\	
 113 &  4:56:44.93 &  $-66:24:29.4$ &$>18.80$&\nodata&$>1.28$ &\nodata&$ 0.77$ & 0.12 \\ 
 129 &  4:56:45.96 &  $-66:24:21.8$ &$>18.80$&\nodata&$>0.95$ &\nodata&$ 0.78$ & 0.14 \\     
 147 &  4:56:47.13 & 	$-66:24:30.5$ &  17.56 & 0.11 &	$0.86 $ & 0.14 & $ 1.22$ & 0.12 \\	
 148 &  4:56:47.17 &  $-66:24:33.1$ &$>18.80$&\nodata&$>0.88$ &\nodata&$ 1.20$ & 0.14 \\
 152 &  4:56:47.33 & 	$-66:24:37.2$ &  17.33 & 0.15 &	$0.17 $ & 0.19 & $ 0.74$ & 0.20 \\	
 157 &  4:56:47.54 & 	$-66:24:33.5$ &  17.05 & 0.05 &	$0.80 $ & 0.07 & $ 1.10$ & 0.07 \\	
 162 &  4:56:47.74 & 	$-66:24:33.8$ &  16.83 & 0.06 &	$0.65 $ & 0.08 & $ 0.58$ & 0.11 \\	
 165 &  4:56:47.88 & 	$-66:24:35.6$ &  16.84 & 0.12 &	$0.12 $ & 0.17 & $ 0.37$ & 0.18 \\	
 169 &  4:56:48.39 & 	$-66:25:21.3$ &  18.55 & 0.09 &	$0.40 $ & 0.12 & $ 0.65$ & 0.11 \\	
 176 &  4:56:48.87 & 	$-66:25:22.2$ &  18.80 & 0.11 &	$0.51 $ & 0.14 & $ 0.91$ & 0.12 \\	
\enddata
\end{deluxetable}

\clearpage

\end{document}